\documentclass{amsart}

\usepackage{amsmath,amssymb}
\usepackage{latexsym}
\usepackage{graphicx}
\usepackage{color}
\usepackage[framemethod=TikZ]{mdframed}
\usepackage{subfigure}
\usepackage{multirow}
\usepackage{mathptmx}
\usepackage{bm}
\usepackage{url}
\usepackage{algorithm}
\usepackage{algorithmic}
\usepackage{rotating}

\usepackage{tikz}
\usetikzlibrary{arrows,positioning,shapes.geometric}

\newcommand{\ud}{\,\mathrm{d}}
\newcommand{\RR}{\mathbb{R}}

\newcommand{\ZZ}{\mathbb{Z}}
\newcommand{\NN}{\mathbb{N}}
\newcommand{\CC}{\mathbb{C}}

\newtheorem{thm}{Theorem}[section]

\newtheorem{Assumption}[thm]{Assumption}

\newcommand{\argmax}{\operatornamewithlimits{argmax}}

\title[single channel fECG by de-shape STFT and nonlocal median]{Extract fetal ECG from single-lead abdominal ECG by de-shape short time Fourier transform and nonlocal median}

\author[L.~Su]{Li~Su}
\address{Research Center for Information Technology Innovation, Academia Sinica}
\email{li.sowaterking@gmail.com}

\author[H.-T.~Wu]{Hau-Tieng~Wu}
\address{Department of Mathematics, University of Toronto}
\email{hauwu@math.toronto.edu}

\begin{document}

\begin{abstract}
The multiple fundamental frequency detection problem and the source separation problem from a single-channel signal containing multiple oscillatory components and a nonstationary noise are both challenging tasks. To extract the fetal electrocardiogram (ECG) from a single-lead maternal abdominal ECG, we face both challenges.
In this paper, we propose a novel method to extract the fetal ECG signal from the single channel maternal abdominal ECG signal, without any additional measurement.
The algorithm is composed of three main ingredients. First, the maternal and fetal heart rates are estimated by the de-shape short time Fourier transform, which is a recently proposed nonlinear time-frequency analysis technique; second, the beat tracking technique is applied to accurately obtain the maternal and fetal R peaks; third, the maternal and fetal ECG waveforms are established by the nonlocal median.  
The algorithm is evaluated on a simulated fetal ECG signal database ({\em fecgsyn} database), and tested on two real databases with the annotation provided by experts ({\em adfecgdb} database and {\em CinC2013} database). In general, the algorithm could be applied to solve other detection and source separation problems, and reconstruct the time-varying wave-shape function of each oscillatory component.
\end{abstract}

\maketitle

\section{Introduction}

Electrocardiograph (ECG) is inarguably the most widely applied measurement to non-invasively study the cardiac activity, since its appearance in 1901 \cite{AlGhatrif_Lindsay:2012}. Its waveform provides a significant amount of clinical information. In addition, the time-varying speed of heart beating, widely understood as the heart rate variability (HRV), has proved to be a portal to our physiological dynamical status. While it has been that widely applied in different scenarios, its application to the intra-uterus fetus is still limited, mainly due to the lack of a direct contact measurement of the fetal ECG (fECG) signal.
Like the adult ECG signal processing, there are two main purposes in the fECG signal processing. First, we want to non-invasively obtain the fetal heart rate, which is intimately related to the fetal distress \cite{Jenkins1989}; second, we would like to analyze the fECG morphology for the sake of diagnosing cardiac problems. However, the fECG morphological analysis is less performed in clinics, except for the ST analysis (STAN) monitor, which detects and alerts the potential risk for fetal hypoxia (see, e.g. \cite{Belfort_Saade:2015} and the citations therein).

There are mainly two types of fECG signals. The first kind of signal is directly recorded through an electrode attached to the fetal skin. For example, the electrode could be attached to the scalp while the cervix dilates during delivery, which is considered invasive. While the recorded signal is of high quality, it can only be recorded during a specific and short period and the instrument is not designed for the long-term monitoring purpose. Also, the infection risk is not negligible. So it is not routinely used in clinics. We call it the {\em direct fECG signal}, and we mention that the STAN monitor depends on the direct fECG signal.
The second kind of signal is recorded from the mother's abdomen, where the sensor is close to the fetus so that the fECG signal is big enough compared with the maternal ECG. The recorded signal is called the {\em abdominal ECG} (aECG), which is composed of the maternal cardiac activity, called the \textit{maternal abdominal ECG (maECG)}, and the fetal cardiac activity, called the {\em indirect fECG signal} (or noninvasive fECG signal). When there is no danger of confusion, we call the indirect fECG signal simply the fECG signal in this paper. An excellent summary of the available measurement techniques and fECG history (as well as several other topics) is provided in \cite{Jenkins1989,Sameni2010}.

While the aECG signal is non-invasive, easy-to-obtain, and suitable for the long-time monitoring purpose, however, from the signal processing viewpoint, it is challenging to obtain the indirect fECG signal from the aECG signal. For example, the fECG signal is always ``contaminated'' by or mixed with the maECG, and the signal-to-noise ratio (SNR) is in general low. These issues challenge the estimation of the fECG and hence the HRV analysis from the aECG signal.
Furthermore, even if the maECG signal could be successfully decoupled from the fECG signal and perfectly denoised, interpreting the morphology of fECG signal is still challenging. This issue originates from the individual variation among subjects, for example, the uterus position and shape, and the fetal size and presentation. So, even if we could standardize the lead system on the mother's abdomen, the application of the fECG waveform is still limited.

The challenge has attracted a lot of attention in the past decades, and several algorithms have been proposed. As is summarized in \cite{Clifford2014}, most methods take the following five steps to study the aECG:
first, pre-process the aECG;
second, estimate the maECG;
third, remove the maECG from the aECG;
fourth, post-process the remainder to obtain the fECG and/or estimate the R peaks and hence the fIHR.
In short, the maECG is removed first so that the fECG could be analyzed from the remainder.
While available algorithms could be classified in different ways based on different criteria, here, for the work presented in this paper, we summarize the existing algorithms based on the number of needed leads, and classify them into two categories  -- one depends on more than one ECG channel and one depends on only one aECG signal.

Most algorithms need multiple aECG channels and/or one maternal thoracic ECG (mtECG) signal, or at least one aECG channel and one mtECG;
for example, 
blind source separation (BSS) \cite{Lathauwer2000,Akhbari2013,DiMariaLiu2014,Varanini2014},
semi-BSS like periodic component analysis ($\pi$CA), or $\pi$Tucker decomposition, which takes the pseudo-periodic structure into account \cite{SameniJutten2008,Haghpanahi2013,Akbari2015},
% NOTE: Akhbari2013, Akbari2015: deterministic BSS based on tensor decomposition.
% NOTE: Haghpanahi2013: obtain a higher SNR fECG by the kurtosis-based short segment merging (polarity detection)
echo state neural network \cite{Behar2014}, least mean square (LMS) \cite{Widrow1975}, recursive least square (RLS) \cite{Behar2014}, and blind adaptive filtering \cite{Graupe2008},
Kalman filter \cite{Sameni2008,Niknazar2013,AndreottiRiedl2014},
% NOTE: echo state neural network Behar2014 is a special adaptive filtering
channel selection approach based on features extracted by different methods, like discrete wavelet transform \cite{Ghaffari2015}, time-adaptive Wiener-filter like filtering \cite{Rodrigues2014},
principal component regression \cite{Lipponen2013},
phase space embedding \cite{KarvounisTsipouras2009},
% NOTE: SameniJutten2008: need the tECG or other ways to get the maternal's R peaks.
to name but a few.

On the other hand, fewer algorithms depend on the single-lead aECG signal; for example,
template subtraction (TS) \cite{VanBemmel1966,Cerutti1986,Martens2007,Ungureanu2007,Behar2014},
and its variation based on singular value decomposition (SVD) or principal component analysis \cite{Kanjilal1997,ChristovSimova2014}, the time-frequency analysis, like wavelet transform, pseudo-smooth Wigner-Ville distribution \cite{Khamene2000,KarvounisTsipouras2007,CastilloMorales2013,Almeida2014} (in practice, three aECG channels are averaged in \cite{KarvounisTsipouras2007}), and S-transform \cite{Lamesgin2015},
sequential total variation \cite{LeeLee2016},
adaptive neuro-fuzzy inference system and extended Kalman filter \cite{PanigrahySahu2015},
particle swarm optimization and extended Kalman smoother \cite{PanigrahyRakshit2015}
state space reconstruction via lag map \cite{Richter1998,Kotas2010},
etc.
We refer the reader with interest to, for example, \cite{Sameni2010,Andreotti2016} for a more detailed review of available methods.

The above-mentioned algorithms all have their own merits and disadvantages; for example, algorithms depending on multiple leads usually provide a more accurate result, but the dependence on multiple leads render it less applicable for the screening and monitoring purpose; on the other hand, the algorithms depending on the single-lead aECG signal usually have lower accuracy, while they could be applied to a wider range of situations.
To simultaneously fulfill the practical need and the accuracy, in this paper, we propose a novel algorithm to extract fetal instantaneous heart rate (fIHR) and the fECG signal from the {\em single-lead aECG signal} from a different viewpoint.
The proposed algorithm combines a recently developed nonlinear time-frequency (TF) analysis called the de-shape short time Fourier transform (de-shape STFT), and the nonlocal median;
the de-shape STFT extracts the maternal instantaneous heart rate (mIHR) from the single-lead aECG, which provides the maternal R peak information. The maECG is then extracted from the aECG by the nonlocal median algorithm. The difference between the aECG and estimated maECG serves as a rough fECG estimate, and the fIHR could be estimated from the rough estimate of fECG by de-shape STFT, and hence the fetal R peaks. The fECG is then extracted by the nonlocal median algorithm. The R peak information could be accurately estimated by the beat tracking algorithm based on the dynamic programming.
While not explicitly used in the algorithm, we mention that our method has the ability to simultaneously obtain the fIHR and mIHR, and hence simultaneously the fECG and mECG.

The novelty and the main difference between our proposed method and the other algorithms based on the single-lead aECG signal are two folds.
First, we use more information hidden in the single-lead aECG signal. Note that the traditional R peak detection algorithms mainly count on the morphological landmarks (fiducial points), like the maximal points representing the R peaks,
or the maximal ``energy'' pattern driven by the QRS complex in the TF domain determined by, for example, the wavelet transform. Then the mIHR and fIHR are obtained by interpolating the estimated R peak locations. 
On the other hand, de-shape STFT allows us to {\em directly} extract the mIHR and fIHR from the single-lead aECG signal, and the fIHR from the rough fECG estimate. We could then utilize the mIHR and fIHR to guide an accurate R peak detection.
Unlike the traditional approach, we simultaneously use the frequency information (the mIHR and fIHR), which reflects the time-varying and nonlinear beat-to-beat relationship, and the morphological landmark information.
Second, based on the nonlinear manifold model, we apply the nonlocal median algorithm \cite{ChaudhurySinger2012,Lin_Minasin_Wu:2016} to extract the maECG and fECG signals -- 
for each cardiac activity candidate, we only consider those aECG segments with a similar pattern, and use the median to estimate the underlying cardiac activity.
Compared with the traditional TS methods \cite{Kanjilal1997}, where the mean, or the mean together with the first few principal components, of consecutive aECG segments containing cardiac activities is considered to be the template of the cardiac activity, the nonlocal median algorithm takes care of the following commonly encountered issues. The fact that the QRST complex morphology is time-varying \cite{MalikFarbom2002} might be overlooked in the traditional TS procedure; the mean of consecutive aECG segments containing cardiac activities is well known to be sensitive to outliers; the TS algorithm is sensitive to the number of principal components and an empirical optimization is needed \cite{ChristovSimova2014}. %The nonlocal median algorithm, on the other hand, could bypass these limitations with theoretical supports. 
In sum, the de-shape STFT is applied to get a better mIHR and fIHR and hence maternal and fetal R peaks, and the nonlocal median is applied to get a better mECG and fECG.

The paper is organized in the following. In Section \ref{Section:BackgroundModel}, we discuss a {\em phenomenological model} for the aECG and the mathematical background for the de-shape STFT and nonlocal median. In Section \ref{Section:MaterialMethod}, the single-lead fECG extraction algorithm is introduced. The material and results are reported in Section \ref{Section:MaterialResult}. In Section \ref{Section:Discussion}, the paper is summarized by a discussion, including limitations and future works.

\section{Mathematical Background}\label{Section:BackgroundModel}

The aECG signal contains at least two components of interest: the component associated with the maternal cardiac activity and the component associated with the fetal cardiac activity, which have different time-varying frequencies and different non-sinusoidal oscillations. These lead to a wide spectrum, so the usual linear signal processing techniques do not work. While it is challenging enough to separate these two components, the problem becomes more challenging considering the influence of different kinds of noise in the measurement. Furthermore, due to the physiological nature of the ECG signal, the non-sinusoidal oscillation is not just impulse-like but also time-varying (see, for example, the nonlinear relationship between QT and RR intervals \cite{MalikFarbom2002}); that is, we will run into the time-varying wave-shape function issue. In this section, we provide a phenomenological model suitable for the aECG signal and an algorithm suitable for analyzing this kind of signal.
Second, we provide a low dimensional and nonlinear geometric model to describe the maternal and fetal cardiac activities. Based on this nonlinear model, the nonlocal median algorithm is introduced to reconstruct the time-varying wave-shape function, and hence extract the mECG and fECG from the aECG.

\subsection{Adaptive non-harmonic model and de-shape short time Fourier transform}\label{Subsection:Model}

To extract the fECG information from the aECG signal, we propose to apply the {\em adaptive non-harmonic model} to model the aECG signal, and apply the de-shape STFT to study the aECG.

\subsubsection{Adaptive non-harmonic model}

We start from introducing the adaptive non-harmonic model.
Take a small enough $0\leq \epsilon<1$, a non-negative sequence $c =\{c(\ell)\}_{\ell=0}^\infty$, $0<C<\infty$ and $N\in\NN$. The set of functions $\mathcal{D}_{\epsilon}^{c,C,N}$ in the space of bounded and continuous functions with continuous first order derivatives, denoted as $C^1(\RR)\cap L^\infty(\RR)$, consists of functions
\begin{align}
x(t)=\frac{1}{2}B_0(t)+\sum_{\ell=1}^\infty B_\ell(t)\cos(2\pi \phi_\ell(t))\label{model:nonlinearRelationship}
\end{align}
satisfying the following three conditions. First, the {\em regularity condition} says that
\begin{align}
&B_\ell\in C^1(\RR)\cap L^\infty(\RR),\quad\phi_\ell\in C^2(\RR)\nonumber
\end{align}
for each $\ell=0,\ldots \infty$. For all $t\in\RR$, $B_1(t)>0$, $B_\ell(t)\geq 0$ for all $\ell=0,2,3,\ldots,\infty$ and $\phi'_\ell(t)>0$ for all $\ell=1,\ldots,\infty$.
Second, the {\em time-varying wave-shape} condition says that for all $t\in\RR$,
\begin{equation}
\left|\frac{\phi'_\ell(t)}{\phi'_1(t)}-\ell\right|\leq \epsilon
\end{equation}
for all $\ell=0,1,\ldots,\infty$,
\begin{equation}
\epsilon<\frac{B_\ell(t)}{B_1(t)}\leq c(\ell)
\end{equation}
for all $\ell=1,\ldots,N$, $\sum_{\ell=N+1}^\infty B_\ell(t)\leq \epsilon B_1(t)$, and $\sum_{\ell=0}^\infty \ell c(\ell)\leq C$.
Third, the {\em slowly varying} condition says that for all $t\in\RR$,
\begin{align}
&|B_\ell'(t)|\leq \epsilon c(\ell)\phi_1'(t),\quad|\phi_\ell''(t)|\leq \epsilon \ell\phi_1'(t), \end{align}
for each $\ell=0,\ldots \infty$, and $\|\phi_1'(t)\|_{L^\infty}<\infty$.

We call a function $x$ in $\mathcal{D}_{\epsilon}^{c,C,N}$ an {\it adaptive non-harmonic (ANH) function}, where the adjective {\em non-harmonic} indicates the possibly non-sinusoidal nature of the oscillation, and the adjective {\em adaptive} indicates the time-varying nature of the frequency, amplitude, and the non-sinusoidal oscillatory pattern.
We call $B_1(t)\cos(2\pi \phi_1(t))$ the {\it fundamental component}, $B_1(t)$ the {\em fundamental amplitude}, $\phi_1$ the {\it phase function}, and $\phi'_1$ the {\it fundamental instantaneous frequency (IF)} of the signal $x$. By a slight abuse of terminology, for $\ell>1$, we call $B_\ell\cos(2\pi\phi_\ell(t))$ the {\it $\ell$-th multiple} of the fundamental component, which we simply call the $\ell$-th multiple if no danger of confusion is possible,
$B_\ell(t)$ the  {\em amplitude of the $\ell$-th multiple}, and $\phi'_\ell$ the {\it IF of the $\ell$-th multiple}, although $\phi_\ell$ might not be an exact integral multiple of $\phi_1$.
Thus, we could view an ANH function as an oscillatory component with the {\it time-varying amplitude, frequency, and wave-shape function}. 

A special case deserves a discussion. When $\beta_\ell:=\frac{B_{\ell}(t)}{B_1(t)}$ are constants for all $\ell=0,1,\ldots,\infty$ and $\phi'_\ell(t)=\ell\phi_1'(t)+\alpha_\ell$ for some $\alpha_\ell\in\RR$ for all $\ell=1,\ldots,\infty$, (\ref{model:nonlinearRelationship}) is reduced to the 
\begin{align}
x(t)=B_1(t)s(\phi_1(t))\label{model:linearRelationship},
\end{align}
where $s$ is a $1$-periodic function with the Fourier coefficients determined by $\beta_\ell$ and $\alpha_\ell$. In this case, clearly the phase function $\phi_1$ is linearly related to the deformation of the non-sinusoidal oscillation, and we say that the wave-shape function is \textit{not} time-varying. On the contrary, for an ANH function, the phase function $\phi_1$ might be nonlinearly related to the deformation of the non-sinusoidal oscillation. If we further assume that $\epsilon=0$, then we obtain the well known harmonic function.

To motivate this model, take the relationship between the RR and QT intervals of an ECG signal as an example.
The nonlinear relationship between the QT interval and the RR interval has been well studied -- for example, the Fridericia's formula (QT interval is proportional to the cubic root of RR interval) or a fully nonlinear depiction \cite{MalikFarbom2002}. Thus, the QRS complex representing the ventricular response is not linearly related to the instantaneous heart rate, and we need a time-varying wave-shape function to model this physiological fact. For more detailed discussion, we refer the reader with interest to \cite{lin2016waveshape}.

In general, a signal might be composed of more than one oscillatory component; for example, the aECG signal is composed of the fECG and maECG. Take a small enough $0<\epsilon<1$ and $d>0$.
We consider the set $\mathcal{D}_{\epsilon,d}\subset C^1(\RR)\cap L^\infty(\RR)$ consisting of superposition of ANH functions; that is,
\begin{equation}\label{Equation:Model:ANH}
f(t)=\sum_{k=1}^{K}f_{k}(t)
\end{equation}
for some finite $K\in\NN$ and 
\begin{equation*}
f_{k}(t)=\sum_{\ell=0}^\infty B_{k,\ell}(t)\cos(2\pi \phi_{k,\ell}(t))\in \mathcal{D}_{\epsilon_k}^{c_{k},C_k,N_k}
\end{equation*}
for some $0\leq \epsilon_k\leq \epsilon$, non-negative sequence $c_{k}=\{c_k(\ell)\}_{\ell=0}^\infty$, $0<C_k<\infty$ and $N_k\in\NN$, where the fundamental IF's of all ANH functions satisfy the following two conditions if $K>1$.
First, the {\em frequency separation} condition says that
\begin{equation}\label{condition_Cepsilon_d}
\phi'_{k,1}(t)-\phi'_{k-1,1}(t)\geq d
\end{equation}
for $k=2,\ldots,K$. Second, the {\em non-multiple} condition says that for each $k=2,\ldots,K$, $\phi'_{k,1}(t)/\phi'_{\ell,1}(t)$ is not an integer for $\ell=1,\ldots,k-1$.
We say that a signal in $\mathcal{D}_{\epsilon,d}$ satisfies the {\em ANH model}.

\subsubsection{Model the maternal abdominal ECG signal by the ANH model}
We now model a recorded aECG signal, $x(t)$, by the ANH model, whichs satisfies
\begin{equation}\label{Equation:Model:aECG}
x(t)=x_m(t)+x_f(t)+n(t)\,,
\end{equation}
where $x_m(t)\in \mathcal{D}_{\epsilon}^{c_m,C_m,N_m}$ for some $c_m,C_m,N_m$ is the maECG signal, $x_f(t)\in \mathcal{D}_{\epsilon}^{c_f,C_f,N_f}$ for some $c_f,C_f,N_f$ is the fECG signal, and $n(t)$ is noise. 
Here, the fundamental IF of $x_m$ (respectively $x_f$) is the mIHR of maECG (respectively fECG).
$n(t)$ includes different kinds of noise, ranging from the baseline wandering, power line interference, maternal electromyographic signal, to uterine contraction, so in general it is not stationary. We consider either smooth varying non-stationary noise model with a smooth and slowly varying covariance function for $n(t)$ to capture the possible heteroscedasticity and autocorrelation inside the noise \cite[Equation (6)]{Chen_Cheng_Wu:2014}, or a more general piecewise locally stationary model \cite[Definition 1]{Zhou2013} to further capture the abrupt change inside the noise structure.

In sum, both $x_m(t)$ and $x_f(t)$ are signals with time-varying IF due to the inevitable HRV with time-varying amplitude and wave-shape function representing the cardiac activity, and the recorded signal might be contaminated by different kinds of inevitable noise.

\subsubsection{de-shape short-time Fourier transform}

There are several challenges of analyzing the aECG signal, including the time-varying amplitude, time-varying frequency, and the time-varying non-sinusoidal oscillation. To deal with this kind of signal, under the ANH model (\ref{Equation:Model:aECG}), we could apply the currently proposed algorithm, the de-shape STFT. The de-shape STFT is a nonlinear TF analysis technique combining the well known STFT and the cepstrum technique commonly applied in the signal processing field, and it is composed of the following four steps. First, note that $x(t)$ is a tempered distribution, so with a chosen window function $h\in\mathcal{S}$, where $\mathcal{S}$ is the Schwartz space, we have
\begin{equation}
V^{(h)}_x(t, \xi) = \int x(\tau) h(\tau-t)e^{-i2\pi \xi (\tau-t)} \ud \tau\,,\label{eq: stft1}
\end{equation}
where $t\in\RR$ indicates time and $\xi\in\RR$ indicates frequency. Clearly, $V^{(h)}_f(t, \xi)\in C^\infty$ is smooth and slowly increasing on both time and frequency axes. Precisely, the frequency function $V^{(h)}_f(t, \cdot)$ for any $t$ (as well as the temporal function $V^{(h)}_f(\cdot, \xi)$ for any $\xi$) and all its derivatives have at most polynomial growth at infinity.
Second, evaluate the {\em short time cepstral transform (STCT)} in order to obtain the fundamental period and its multiples:
\begin{equation}
C^{(h,\gamma)}_x(t, q) := \int |V^{(h)}_x(t, \xi)|^\gamma e^{-i2\pi q \xi} \ud \xi,
\label{eq: rceps1}
\end{equation}
where $\gamma>0$ is sufficiently small and $q\in\RR$ is called the quefency (its unit is second or any feasible unit in the time domain).  Since $V^{(h)}_f(t, \xi)\in C^\infty$ is smooth and slowly increasing, $|V^{(h)}_f(t, \cdot)|^\gamma$ is continuous and slowly increasing. Hence its Fourier transform is well-defined in the distribution sense.
Third, evaluate the {\em inverse short time cepstral transform (iSTCT)}, which is defined on $\RR\times \RR^+$ as
\begin{equation}
U_x^{(h,\gamma)}(t,\xi):=C_x^{(h,\gamma)}(t,1/\xi),
\end{equation}
where $\xi>0$ (its unit is Hz or any feasible unit in the frequency domain) and $U_x^{(h,\gamma)}(t,\cdot)$ is in general a distribution. Precisely, the map $\xi\to 1/\xi$ from $(0,\infty)$ to $(0,\infty)$ is open and its differentiation is surjective on $(0,\infty)$, so for a distribution $C_x^{(h,\gamma)}(t,\cdot)$, for any $t$, defined on $(0,\infty)$, $C_x^{(h,\gamma)}(t,1/\xi)$ is well-defined \cite[Theorem 6.1.2]{Hormander:1990}.
Fourth, we remove all the multiples by evaluating the de-shape STFT, which is defined on $\RR\times \RR^+$ as
\begin{equation}
\label{eq:W}
W^{(h,\gamma)}_x(t, \xi) := V^{(h)}_x(t,\xi)U^{(h,\gamma)}_x(t, \xi),
\end{equation}
where $\xi>0$ is interpreted as frequency. Note that the de-shape STFT is well-defined as a distribution since $V^{(h)}_f(t,\xi)$ is a $C^\infty$ function and $U^{(h,\gamma)}_f(t, \xi)$ is a distribution.

The main motivation of the de-shape STFT is to decouple the IF and the non-sinusoidal wave-shape function. Due to the non-sinusoidal oscillation, at each time $t$, in STFT we could see not only the fundamental frequency but also its multiples. The existence of multiples, when there are more than one component, interfere with each other and mask the true information we have interest. Thus, the notion of cepstrum is applied to obtain the fundamental period information of the signal. Note that the fundamental period and its multiples, after the inversion, become the fundamental frequency and its divisions, and hence the common ingredient between iSTCT and STFT is the fundamental frequency. Thus, after a direct element-wise product, only the fundamental frequency is preserved in the de-shape STFT. This approach could be viewed as a ``nonlinear filter'' technique, which uses the ``dual information'' of the spectrum (the cepstrum), as a mask to remove the irrelevant information (the wave-shape function) and keep the relevant information (the IF).  

It has been shown in \cite[Theorem 3.6]{lin2016waveshape} that under the ANH model (\ref{Equation:Model:ANH}), the de-shape STFT could extract the fundamental IF of each component, where all multiple IF's are suppressed. In other words, the time-varying wave-shape information of maECG and fECG is decoupled from the IF and AM of maECG and fECG in the TF representation.
The implementation of the de-shape STFT in the discrete-time domain will be discussed in Section \ref{Section:MaterialMethod}. 

\subsection{Nonlocal median}

Nonlocal median algorithm \cite{ChaudhurySinger2012,Lin_Minasin_Wu:2016} is a variation of the well-known nonlocal mean algorithm in the image processing field, mainly for the purpose of image denoising \cite{Buades_Coll_Morel:2005,Singer_Shkolnisky_Nadler:2009}. After obtaining the fundamental IF of each component, we apply the nonlocal median algorithm to recover the time-varying wave-shape function. In this study, it allows us to extract the maECG and fECG from the aECG signal.

Without loss of generality, we assume that maECG and fECG are both with the Rs pattern in the aECG so that we could discuss the R peaks. The discussion holds for the S peaks if either maECG or fECG has the rS pattern.
Take the maECG into account. Suppose the $k$-th cardiac activity, which could be normal or ectopic, without any other pathological arrhythmia, in the maECG starts at $s_m^{(k)}\in\RR$ and ends at $e_m^{(k)}\in\RR$, and suppose the $k$-th R peak is located at $r_m^{(k)}\in (s_k,e_k)$. Similarly, we could define $s_f^{(k)},e_f^{(k)}$ and $r_f^{(k)}$ for the $k$-th cardiac activity in the fECG. By the physiological property of the cardiac activity, we know that $e_m^{(k)}<s_m^{(k+1)}$ and $e_f^{(k)}<s_f^{(k+1)}$ for all $k$, where $(e_m^{(k)},s_m^{(k+1)})$ and $(e_f^{(k)},s_f^{(k+1)})$ are periods where the maECG and fECG are isoelectric respectively (known as the TP interval). Thus, the ANH function $x_m$ in (\ref{Equation:Model:aECG}) could be written as
\begin{equation}
x_m(t)\,=\sum_{k\in\ZZ}x_m(t)\chi_{[s_m^{(k)},e_m^{(k)}]},\label{Model:LinearRelationship}
\end{equation}
where $\chi_{[s_m^{(k)},e_m^{(k)})}$ is an indicator function defined on $[s_m^{(k)},e_m^{(k)}]$. Note that due to the time-varying nature of the non-sinusoidal oscillation, the amplitude of the fundamental component and each multiple in the ANH function is nonlinearly related to the waveform morphology. It is well known that electrophysiologically, the fECG and maECG are similar, except the heart rate \cite{Sameni2010}, so the above discussion could be carried over to the fECG, $x_f$. Consider $l_m=\max_k\{r_m^{(k)}-s_m^{(k)}\}>0$ and $r_m=\max_k\{e_m^{(k)}-r_m^{(k)}\}>0$, and denote the $k$-th abdominal cardiac activity mixture as
\begin{equation}
x_m^{(k)}:[0,l_m+r_m]\to \RR,
\end{equation}
where $x_m^{(k)}(t)=x(r_m^{(k)}+t-l_m)$ when $t\in [l_m-(r_m^{(k)}-s_m^{(k)}),l_m+(e_m^{(k)}-r_m^{(k)})]$ and $x_m^{(k)}(t)=0$ otherwise.
Clearly, $x_m^{(k)}$ is the $k$-th aECG segment containing the $k$-th maternal cardiac activities, denoted as
\begin{equation}
x^{(k)}_{m,m}:[0,l_m+r_m]\to \RR,
\end{equation}
where $x_{m,m}^{(k)}(t)=x_m(r_m^{(k)}+t-l_m)$ when $t\in [l_m-(r_m^{(k)}-s_m^{(k)}),l_m+(e_m^{(k)}-r_m^{(k)})]$ and $x_{m,m}^{(k)}(t)=0$ otherwise, and several fetal heart beats, since normally the fetal heart rate is higher.
Denote
\begin{equation}
\mathcal{X}_m:=\{x_m^{(k)}\}_{k\in\NN}\subset C^1([0,l_m+r_m])
\end{equation}
to be collection of aECG segments  from $x(t)$ and
\begin{equation}
\mathcal{X}_{m,m}:=\{x_{m,m}^{(k)}\}_{k\in\NN}\subset C^1([0,l_m+r_m])
\end{equation}
to be collection of maECG segments from the underlying maternal cardiac activities. We could thus view $\mathcal{X}_m$ as a ``noisy'' collection of maECG segments $\mathcal{X}_{m,m}$, and the mission is to recover $\mathcal{X}_{m,m}$ from $\mathcal{X}_m$.
Similarly, we could define $\mathcal{X}_{f}=\{x_f^{(k)}\}_{k\in\NN}\subset C^1([0,l_f+r_f])$ and $\mathcal{X}_{f,f}=\{x_{f,f}^{(k)}\}_{k\in\NN}\subset C^1([0,l_f+r_f])$, where $l_f:=\max_k\{r_f^{(k)}-s_f^{(k)}\}>0$, $r_f:=\max_k\{e_f^{(k)}-r_f^{(k)}\}>0$, $x_f^{(k)}(t)=x(r_f^{(k)}+t-l_f)$ when $t\in [l_f-(r_f^{(k)}-s_f^{(k)}),l_f+(e_f^{(k)}-r_f^{(k)})]$ and $x_f^{(k)}(t)=0$ otherwise, and $x_{f,f}^{(k)}(t)=x_f(r_f^{(k)}+t-l_f)$ when $t\in [l_f-(r_f^{(k)}-s_f^{(k)}),l_f+(e_f^{(k)}-r_f^{(k)})]$ and $x_{f,f}^{(k)}(t)=0$ otherwise.

Physiologically, it is well known that while the underlying mechanism leading to the cardiac activities might be complicated \cite{Andreotti2016}, phenomenologically they are similar from beat to beat. There are two dominant parameters that quantify the similarity between cardiac activities -- the scaling and the dilation, where the scaling reflects the respiratory activity and the dilation reflects the nonlinear relationship between the RR interval and QT interval. Also, the waveform representing the cardiac activity should be bounded and with a bounded differentiation. This fact could be summarized in the following:
\begin{Assumption}\label{Assumption:LowDimensionalManifold}
The dataset $\mathcal{X}_{m,m}$ is sampled from a random vector $V_m$, where $V_m$ has the range supported on a bounded set inside $C^1([0,l_m+r_m])$ with a low dimensional structure. To simplify the qualitative description ``low dimensional structure'', we assume a low dimensional smooth and compact manifold to quantify the range of $V_m$.
\end{Assumption}
Note that due to the fact that the maECG amplitude and frequency are both time-varying, under this model, two consecutive maternal cardiac activities might be far away in the manifold.

Similarly, we have the same assumption for the fetal cardiac activity; that is, the set $\mathcal{X}_{f,f}$ is sampled from a random vector $V_f$, with the range supported on a bounded set inside $C^1([0,l_f+r_f])$ with a low dimensional structure. A critical assumption we need to apply the nonlocal median algorithm is the following
\begin{Assumption}\label{Assumption:independence}
The random vectors $V_m$ and $V_f$ are independent.
\end{Assumption}
This assumption essentially says that for different $x^{(k)}$, while the maternal cardiac activities are similar, the fetal cardiac activities sit in random positions.

With the above setup and assumptions, we could now introduce the nonlocal median algorithm. For each $x_m^{(k)}\in \mathcal{X}_m$, find its $K$ nearest neighbors with the $L^2$ norm, where $K\in\NN$ is chosen by the user. Precisely, by ranking $d_{k,l}:=\|x_m^{(k)}-x_m^{(l)}\|_{L^2}$ in the ascending order, we have the set $\mathcal{N}_m^{(k)}$ containing the first $K$ neighbors of $x_m^{(k)}$ in $\mathcal{X}_m$ with the smallest $L^2$ norm. Note that we could also consider the correlation or other more sophisticated metrics, but to keep the discussion simple, we focus on the $L^2$ norm in this paper. Then, the $k$-th maternal cardiac activity is estimated by
\begin{equation}
\tilde{x}_m^{(k)}(t):=\text{median} \{x_m^{(j)}(t)|\, x_m^{(j)}\in \mathcal{N}_m^{(k)}\}
\end{equation}
for all $t\in [0,l_m+r_m]$. Based on Assumption \ref{Assumption:LowDimensionalManifold}, the neighbors we find have similar maternal cardiac activities while they are not neighbors in the temporal axis. On the other hand, it is well known that the median is robust to outliers. Note that under Assumption \ref{Assumption:independence}, for different $x^{(k)}$, the fetal cardiac activities sit in different positions in the support of $x^{(k)}$, and note that the ECG morphology associated with the ventricular activity is spiky, so the fetal cardiac activity occupies a small portion of the interval $[0,l_m+r_m]$. As a result, the median value will faithfully reflect the maternal cardiac activity at $t$, and hence the recovery of the maECG.

The above procedure could be applied to extract the fECG signal from the aECG signal, if we reverse the role of the maECG and fECG, and consider $\mathcal{X}_{f}$ from $\mathcal{X}_{f,f}$.

\section{Methodology}\label{Section:MaterialMethod}

In this section, we introduce our single-lead fECG extraction algorithm, and provide a summary of the used statistics for the evaluation. %The algorithm is implemented in Matlab, and each component could be found in the author's website.

\subsection{Single-channel fetal ECG extraction algorithm}

We now introduce our algorithm to extract the fECG signal from the aECG signal. The aECG signal $x(t)$ is sampled with the sampling rate $f_s\in\mathbb{N}$ during the interval from the $0$-th second to the $T$-th second, where $T>0$. Denote
\begin{equation}
\mathbf{x}_0:=[x(0),x(1/f_s),\ldots,x(\lfloor T/f_s\rfloor)]^T\in \mathbb{R}^{N},
\end{equation}
where $N=\lfloor T/f_s\rfloor+1$, to be the collected aECG signal. Thus we have
\begin{equation}
\mathbf{x}_0=\mathbf{x}_{m}+\mathbf{x}_{f}+\mathbf{n}\,,
\end{equation}
where $\mathbf{x}_{m}$, $\mathbf{x}_{f}$, and $\mathbf{n}$ are the discretized maECG signal, fECG signal and noise. The algorithm is summarized in the flowchart in Figure \ref{FlowChart}. Below we detail the algorithm step by step.

\tikzstyle{line} = [draw, -latex']
\tikzstyle{arrow} = [thick,->,>=stealth]

\begin{figure}[h!]
   \begin{tikzpicture}[>=latex']
        \tikzset{block/.style= {draw, rectangle, align=center,minimum width=2cm,minimum height=.7cm,line width=0.3mm},
        rblock/.style={draw, shape=rectangle,rounded corners=1.5em,align=center,minimum width=2cm,minimum height=.7cm},
        input/.style={ % requires library shapes.geometric
        draw,
        trapezium,
        trapezium left angle=60,
        trapezium right angle=120,
        minimum width=2cm,
        align=center,
        minimum height=.7cm
    },
        }

        \node [rblock]  (start) {Input \\\textit{the signal lead aECG $\mathbf{x}_0$}\\\includegraphics[width=.4\textwidth]{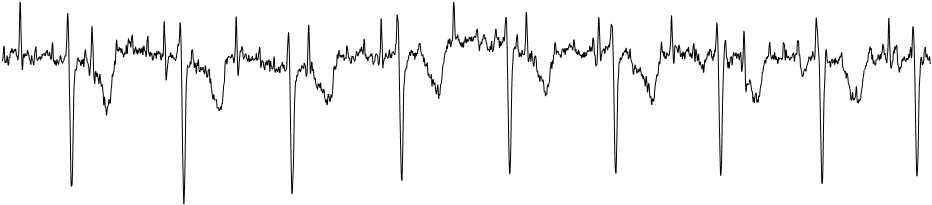}};
        \node [block, right =.5cm of start] (prep) {preprocessing\\\textit{get preprocessed aECG $\mathbf{x}$}};

        \node [block, below right =.5cm and -6cm of start] (dsSTFT) {de-shape STFT\\ \textit{get TFR of $\mathbf{x}$}\\\includegraphics[width=.35\textwidth]{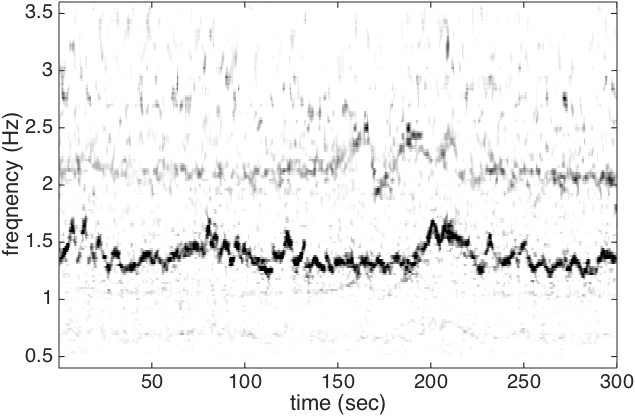}};
        \node [block, right =.5cm of dsSTFT] (mIHR) {Curve extraction\\\textit{get mIHR}\textit{ (and optionally fIHR)}\\\includegraphics[width=.35\textwidth]{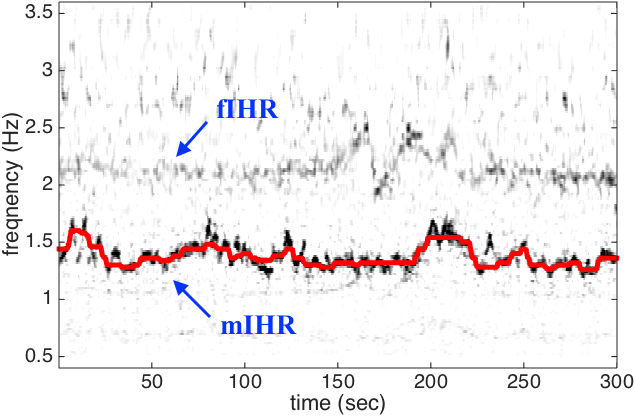}};

        \node [block, below right =1.2cm and -5cm of dsSTFT] (BT) {Beat tracking \\ \textit{get maternal R peaks}\\\includegraphics[width=.4\textwidth]{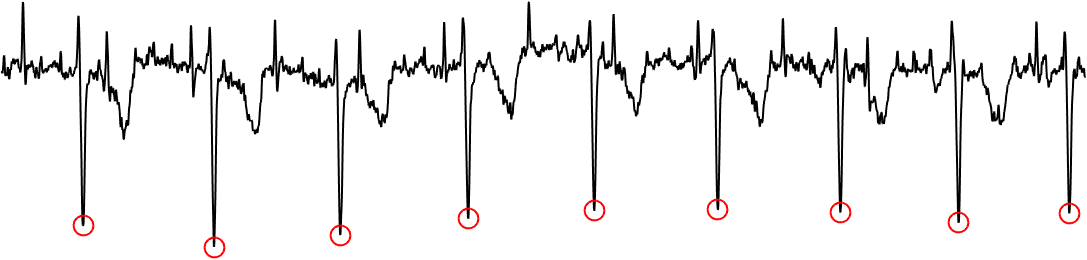}};
        \node [block, right =.5cm of BT] (NLM1) {nonlocal median \\ \textit{get maECG $\tilde{\mathbf{x}}_m$ and} \textit{rough fECG $\tilde{\mathbf{x}}_{f,0}:=\mathbf{x}-\tilde{\mathbf{x}}_m$}\\\includegraphics[width=.4\textwidth]{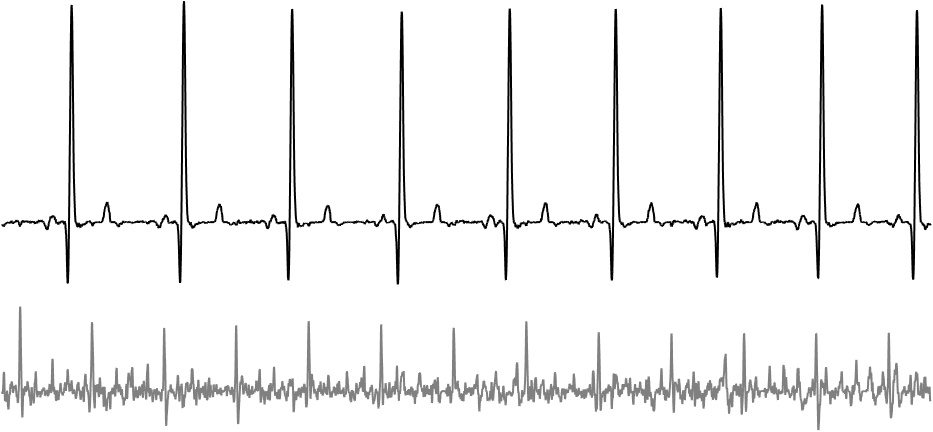}};
%        \node [block, right =.6cm of NLM1] (NLM2) {nonlocal median \\ \textit{get fECG}};

        \node [block, below right =1.2cm and -5cm of BT] (dsSTFT2) {de-shape STFT\\ \textit{get TFR of $\tilde{\mathbf{x}}_{f,0}$}\\\includegraphics[width=.35\textwidth]{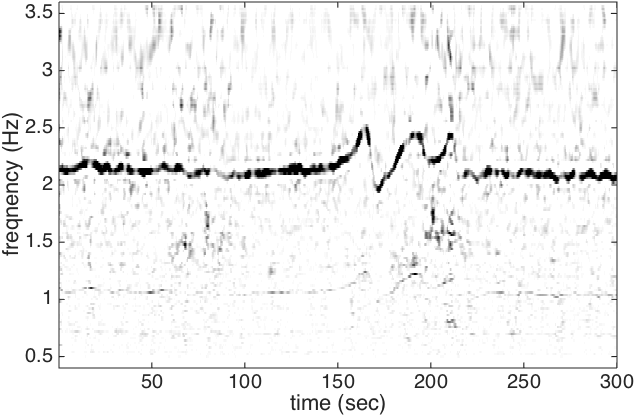}};
       \node [block, right =.5cm of dsSTFT2] (fIHR) {Curve extraction\\\textit{get fIHR}\\\includegraphics[width=.35\textwidth]{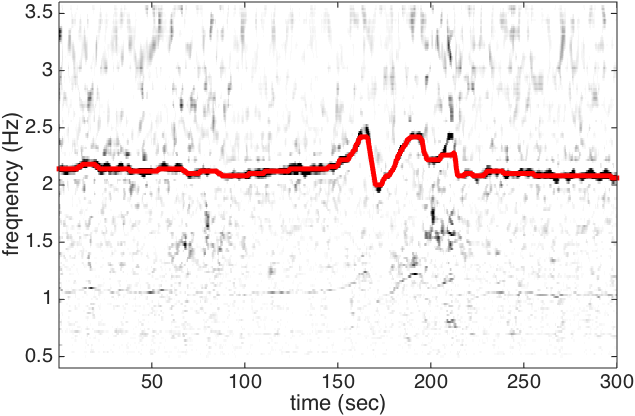}};

        \node [block, below right =.8cm and -5cm of dsSTFT2] (BT2) {Beat tracking \\ \textit{get fetal R peaks}\\\includegraphics[width=.4\textwidth]{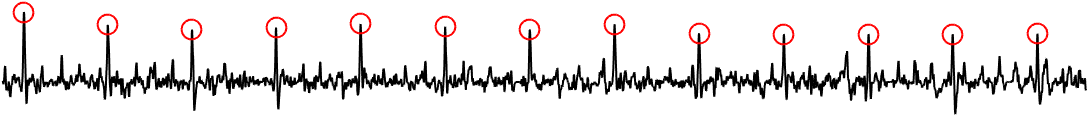}};
        \node [block, right =.5cm of BT2] (NLM2) {nonlocal median \\ \textit{get fECG $\tilde{\mathbf{x}}_f$}\\\includegraphics[width=.4\textwidth]{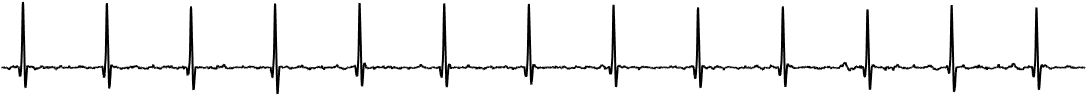}};
         \node [rblock, right =.5cm of NLM2] (Output) {Output \\ \textit{fIHR},\\\textit{fetal R peaks}, \\\textit{and fECG}};

        \node [coordinate, below right =.7cm and .5cm of prep] (right1) {};  %% Coordinate on right and middle
        \node [coordinate, above left =0.3cm and .5cm of dsSTFT] (left1) {};  %% Coordinate on left and middle

        \node [coordinate, below right =.2cm and .5cm of mIHR] (right11) {};  %% Coordinate on right and middle
        \node [coordinate, above left =1cm and .5cm of BT] (left11) {};  %% Coordinate on left and middle

        \node [coordinate, below right =.2cm and .5cm of NLM1] (right2) {};  %% Coordinate on right and middle
        \node [coordinate, above left =.25cm and .5cm of dsSTFT2] (left2) {};  %% Coordinate on

       \node [coordinate, below right =.3cm and .5cm of fIHR] (right22) {};  %% Coordinate on right and middle
        \node [coordinate, above left =.5cm and .5cm of BT2] (left22) {};  %% Coordinate on

        %\node [coordinate, below right =.3cm and 1cm of fIHR] (bottom1) {};  %% Coordinate on right and middle
        %\node [coordinate, above left =.3cm and 1cm of Output] (top1) {};  %% Coordinate on
%% paths
        \path[draw,->]
                  (prep.east) -| (right1) -- (left1) |- (dsSTFT);
         \path[draw,->]
            (mIHR.east) -| (right11) -- (left11) |- (BT);
           \path[draw,->]
         (NLM1.east) -| (right2) --node[anchor=south] {}(left2) |- (dsSTFT2);
            \path[draw,->]
         (fIHR) -| (right22) -- (left22) |-  (BT2);

          \draw[arrow]   (start.east)  --(prep.west);
          \draw[arrow]   (dsSTFT.east)  --(mIHR.west);
          \draw[arrow]   (BT.east)  --(NLM1.west);
          \draw[arrow]   (dsSTFT2.east)  --(fIHR.west);
          \draw[arrow]   (BT2.east)  --(NLM2.west);
          \draw[arrow]   (NLM2.east)  --(Output.west);
    \end{tikzpicture}
    \caption{\label{FlowChart}The flow chart of the proposed algorithm for extracting fECG from the signal lead aECG signal. The shown aECG signal is of 6 seconds long.}
\end{figure}
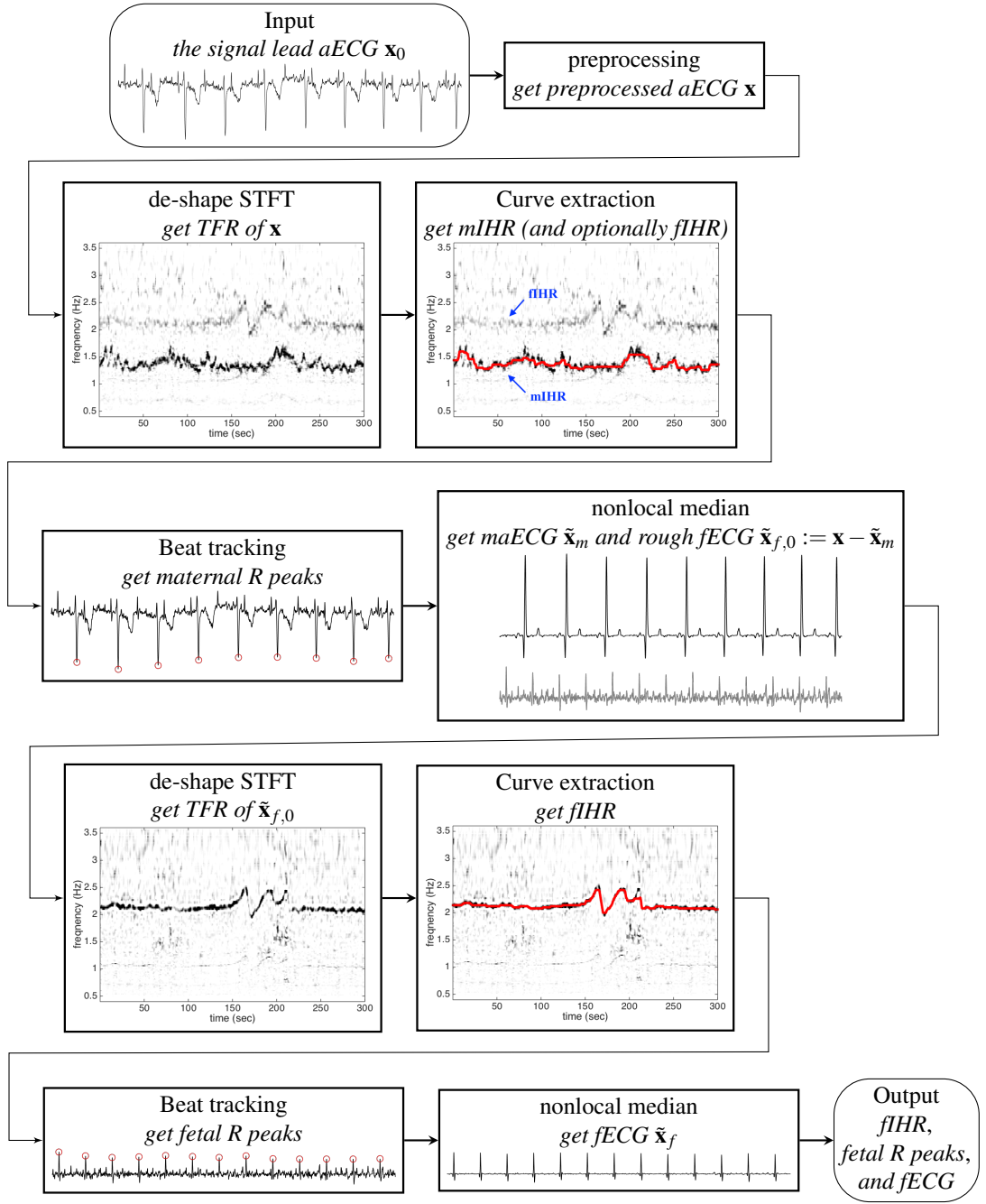

\subsubsection{Step 0: Preprocessing}

The first step of the proposed algorithm, as most other algorithms, is signal preprocessing.  Among all the analyses, we apply the following steps. First, we remove the baseline wandering by the median filter with the window size $100$ ms. 
%Then, we follow the suggestion in \cite{Andreotti2016} and apply a low-pass filtering with the cut-off frequency of 100 Hz to reduce the noise influence.
%
If needed, the power-line interference is suppressed by a zero-phase notch filter at $50$ or $60$ Hz. 
To preserve the non-stationary nature of the signal, we do not apply any other linear filtering technique to the signal.
If the signal is sampled at the frequency lower than $1000$ Hz, to enhance the R peak alignment needed in the nonlocal median step, the signal is upsampled to $1000$ Hz \cite{LagunaSornmo2000}. To simplify the notation, we use the same notation $f_s$ and $N$ to denote the resulting sampling rate and the resulting number of sampling points, and denote the resulting signal as $\mathbf{x}\in\RR^N$.

\subsubsection{Step 1: Run de-shape STFT on $\mathbf{x}$ to estimate the maternal instantaneous heart rate}

We apply the de-shape STFT algorithm to extract the IF information of the maternal and fetal cardiac activities from the preprocessed aECG signal $\mathbf{x}$.
Fix the frequency resolution of the STFT by $\frac{f_s}{2M}$, where $M\in\NN$ is the number of discretization points in the frequency axis, and the quefency resolution by $\frac{M}{f_sM'}$, where $M'\in\NN$ is the number of discretization points in the quefency axis. The numerical implementation of the de-shape STFT algorithm is summarized in Algorithm \ref{alg:deShapeSST}. Denote $\mathbf{W}_{\mathbf{x}}\in\mathbb{C}^{N,M}$ to be the de-shape STFT of $\mathbf{x}$.
\begin{algorithm}[h!]
\begin{algorithmic}
\STATE[INPUT] the aECG signal $\mathbf{x}\in \RR^N$; a discretized window function $\mathbf{h}\in \RR^{2k+1}$, $k\in\NN$; a sufficiently small $\gamma>0$; an upsample factor $\alpha\in\NN$. 
\STATE[STEP 1-1] Evaluate the STFT of $\mathbf{x}$, denoted as $\mathbf{V}_{\mathbf{x}}\in\CC^{N,2M}$, where $M\in\NN$ is related to the chosen frequency resolution, by
\begin{equation}
\mathbf{V}_{\mathbf{x}}(n,m) = \frac{1}{f_s}\sum_{l=-k}^{k}\mathbf{x}(n+l) \mathbf{h}(l+k+1)e^{-i\pi n(m-M)/M}\,,\nonumber
\end{equation}
where $m=1,\ldots,2M$ and we pad $\mathbf{x}$ by $0$ so that $\mathbf{x}(l)=0$ when $l<1$ and $l>N$.

\STATE[STEP 1-2] Evaluate the STCT of $\mathbf{x}$, denoted as $\mathbf{C}_{\mathbf{x}}\in\mathbb{C}^{N,2M}$ by:
\begin{equation}
\mathbf{C}_{\mathbf{x}}(n,m') := \frac{f_s}{2M}\sum_{m=1}^{2M} |\mathbf{V}_{\mathbf{x}}(n,m)|^\gamma e^{-i\pi m' (m-M)/M}, \nonumber
\end{equation}
where $m'=1,\ldots,M$.

\STATE[STEP 1-3] Upsample the positive quefency axis of $\mathbf{C}_{\mathbf{x}}$ to be $\alpha$-time finer; that is, construct a new matrix $\tilde{\mathbf{C}}_{\mathbf{x}}\in\mathbb{C}^{N,\alpha M}$, where 
\begin{equation}
\tilde{\mathbf{C}}_{\mathbf{x}}(n,m'') := \mathbf{C}_{\mathbf{x}}\left(n,M+\frac{m''}{\alpha}\right)\nonumber
\end{equation}
for all $n=1,\ldots,N$ and $m''=1,\ldots,2\alpha M$.

\STATE[STEP 1-4] Evaluate the iSTCT of $\mathbf{x}$, denoted as $\mathbf{U}_{\mathbf{x}}\in\CC^{N,M}$, by:
\begin{equation}
\mathbf{U}_{\mathbf{x}}(n,m):=\sum^{\lceil m+1/2\rceil}_{1/m''=\lceil m-1/2\rceil}\tilde{\mathbf{C}}_{\mathbf{x}}\left(n,\frac{1}{m''}\right),\nonumber
\end{equation}
where $n=1,\ldots,N$ and $m=1,\ldots,M$.

\STATE[STEP 1-5] Evaluate the de-shape STFT of $\mathbf{x}$, denoted as $\mathbf{W}_{\mathbf{x}}\in\mathbb{C}^{N,M}$, by:
\begin{equation}
\label{eq:W}
\mathbf{W}_{\mathbf{x}}(n,m) := \mathbf{V}_{\mathbf{x}}(n,m)\mathbf{U}_{\mathbf{x}}(n,m),\nonumber
\end{equation}
where $m=1,\ldots,N$ and $n=1,\ldots,M$.

\STATE[OUTPUT] the de-shape STFT of $\mathbf{x}$, $\mathbf{W}_{\mathbf{x}}$, for the postprocessing.
\end{algorithmic}
\caption{de-shape STFT algorithm.}
\label{alg:deShapeSST}
\end{algorithm}

It has been systematically reported in \cite{lin2016waveshape} that the fundamental frequencies of the maECG and fECG are represented as dominant curves in the TFR. This fact allows us to extract the salient fundamental IF information of each oscillatory component \cite{lin2016waveshape}.
While the feature inside TFR determined by the de-shape STFT is suitable for several IF tracking methods, including dynamic programming (DP), dynamic Bayesian networks, adaptive filters, and others, to simplify the discussion, we apply the simple DP curve extraction algorithm \cite{Chen_Cheng_Wu:2014} to track the peaks and estimate the IFs based on one assumption that the maECG is strong than the fECG. The detailed procedure is as follows:

\begin{enumerate}
\item Performing DP on $\mathbf{W}_{\mathbf{x}}$ to extract the most dominant component. The extracted curve, when adjusted with the frequency resolution, denoted as $\eta_m\in\mathbb{R}^N$, represents the mIHR. Precisely, suppose at time $n$, the most dominant component is located at $\mathbf{W}_{\mathbf{x}}(n,j_n)$, then $\eta_m(n)=\frac{j_nf_s}{2M}$.
\item (Optional step 1: simultaneously estimate the fIHR) For every time $n$, multiply $\mathbf{W}(n, j)$ by $\theta_m$, where $j\in\{\eta_m(n)-N_\Delta,\eta_m(n)-N_\Delta+1,\ldots,\eta_m(n)+N_\Delta-1,\eta_m(n)+N_\Delta\}$ and $N_\Delta\in\NN$ and $0\leq \theta_m<1$ are chosen by the user, to suppress the maternal cardiac activity. This procedure makes the fECG be the predominant component in $\mathbf{W}_{\mathbf{x}}$. Then, performing the curve extraction on $\mathbf{W}_{\mathbf{x}}$ again to extract the fIHR, denoted as $\eta_f\in\mathbb{R}^N$. In practice, we could choose $N_\Delta$ so that $\frac{N_\Delta f_s}{2M}=0.1$ Hz and $\theta_m=10^{-4}$. This optional step is not carried out in this paper.
\end{enumerate}

The temporal complexity of evaluating the de-shape STFT is $O(NM\log M)$. Indeed, the evaluation of STFT is local in nature, and it depends on the chosen window and the frequency resolution.

\subsubsection{Step 2: Obtain the maternal R peaks by beat tracking and dynamic programming}

Note that although theoretically the IHR is related to the R peak to R peak interval (RRI) time series, in practice there is a discrepancy due to the time-varying nature of the wave-shape function. To obtain the exact R peak location, we apply the beat tracking technique, which has been well studied in music signal analysis\footnote{The {\em beat tracking} problem in music refers to finding the instants of beats in music by analyzing the accents of music signals (e.g., drum hits).}.

Define the estimated maternal RRI as $\delta_m(n):=1/\eta_m(n)$, which is the inverse of the mIHR. Our goal is to find an strictly increasing sequence $B_m=\{b_i\}_{i=1}^{M_m}$ of length $M_m$, where $M_m\in \NN$ and $0< b_i\leq  N$ so that $b_i$ is the index of $i$-th maternal R peak position and $(b_i-b_{i-1})/f_s$ is close to the estimated maternal RRI at time $b_i/f_s$, $\delta_m(b_i)$. We call $B_m$ the \textit{maternal beat sequence}. This problem is understood as the beat tracking problem, and it is formulated as the following optimization problem \cite{ellis2007beat}:
\begin{equation}
\tilde{B}_m=\argmax_{B_m} \Big[\sum^{M_m}_{i=1}\mathbf{x}(b_i) + \lambda_{\texttt{BT}} \sum^{M_m}_{i=2} P(b_i, b_{i-1})\Big],
\label{eq:dpbeat}
\end{equation}
where $P(b_i, b_{i-1}):=-\left(\log_2\left(\frac{(b_i-b_{i-1})/f_s}{\delta_m(b_i)}\right)\right)^2$ and $\lambda_{\texttt{BT}}\geq 0$ is the penalty term determined by the user, which balances the influence of R peak amplitude and the estimated maternal RRI. Notice that \cite{ellis2007beat} assumes $\delta_m(n)$ (the inter-beat interval in the music clip; in our case, the IHR) being a constant function, which is clearly not suitable for an oscillatory signal with time-varying frequency. To address this issue, we modify the original formulation in \cite{ellis2007beat} by allowing a time-varying function $1/\eta_m(n)$.
Note that the first term in the objective function in (\ref{eq:dpbeat}) reaches its maximal value when $b_i$ matches the R-peak location, and the second term penalizes the discrepancy between the estimated IHR and the IHR determined by the true RR interval, with the maximal value zero. 
The resulting beat sequence $\tilde{B}_m=\{\tilde{b}_i\}_{i=1}^{M_m}$ provides an estimate of R peaks location; that is, $\tilde{b}_i$ is an estimate of the $i$-th R peak location.
In our experiments, setting $\lambda_{\texttt{BT}}$ between 20 and 50 turns out to yield a suitable tradeoff, and the result is not sensitive to $\lambda_{\texttt{BT}}$.

While the number of possible beat sequences in $\mathbf{x}$ grows exponentially as $N$ grows, the optimization problem (\ref{eq:dpbeat}) can be solved effectively by the DP. The main idea leading to the DP is that the objective function in (\ref{eq:dpbeat}) is an accumulation in time. Thus, we could divide the problem into a set of optimization subproblems, each of which optimizes the objective function up to a prescribed time step, and the solution is the combination of the optima at every time step. This is implemented by introducing two additional vectors $\mathbf{C}\in\RR^N $ and $\mathbf{D}\in\RR^N $, where $\mathbf{C}(n)$ records the maximum of the objective function accumulated from 1 to $n$, where $n=1,\ldots,N$, and $\mathbf{D}(n)$ records the estimated beat position yielding this maximum at the time step of $n-1$. $\mathbf{D}$ allows us to trace previous beat positions, and this step is called the {\em backtracking}. The whole procedure is sketched in Algorithm \ref{alg:DPecg}. More details of this method can be found in \cite{ellis2007beat}, and the source code is available in \url{http://labrosa.ee.columbia.edu/projects/beattrack/}.

\begin{algorithm}[h!]
\begin{algorithmic}
\STATE[INPUT] the aECG signal $\mathbf{x}\in \RR^N$; an estimated sequence of RR intervals $\delta_m$; an weighted parameter $\lambda_{\texttt{BT}}$;
\STATE[STEP 2-1] Initialize $\mathbf{C}(0)=0$, $\mathbf{D}(0)=0$

\STATE[STEP 2-2] Accumulate the objective function and store the result

\FOR{$i = 1$ \TO $N$}
\STATE $\mathbf{C}(n) = \mathbf{x}(n)+\max_{m=1,\ldots,n-1}[\mathbf{C}(m)+\lambda_{\texttt{BT}} P(n,m)]$;
\STATE $\mathbf{D}(n) = \argmax_{m=1,\ldots,n-1}[\mathbf{C}(m)+\lambda_{\texttt{BT}} P(n,m)]$;
\ENDFOR

\STATE[STEP 2-3] Backtracking
\STATE $a_1 = \argmax_{n} \mathbf{C}(n)$;
\WHILE {$\mathbf{D}(a_l)>0$}
\STATE $l \leftarrow l+1$; $a_l = \mathbf{D}(a_{l-1})$;
\ENDWHILE

\STATE[OUTPUT] the optimal beat sequence $(b_1, \cdots, b_{M_m})$, where $b_j=a_{M_m-j+1}$, $j=1,\ldots,M_m$.
\end{algorithmic}
\caption{Beat tracking by dynamic programming.}
\label{alg:DPecg}
\end{algorithm}

\subsubsection{Step 3: Estimate the maECG morphology by the nonlocal median}

With the maternal R peak location, we could extract $\mathbf{x}_m$ from the aECG signal.
Based on the physiological knowledge, we choose $L_m,R_m\in\NN$ so that $[\frac{\tilde{b}_i-L_m}{f_s},\frac{\tilde{b}_i+R_m}{f_s}]$, where $i=1,\ldots,M_m$, is long enough to cover the $i$-th maternal cardiac activity. Define the aECG segment $\mathbf{x}_i \in \mathbb{R}^{L_m+R_m+1}$ associated with the $i$-th maternal cardiac activity, where $i=1,\ldots, M_m$ and
\begin{align}
\mathbf{x}^{(i)} := \big[ \mathbf{x}(\tilde{b}_i-L_m) \dots \mathbf{x}(\tilde{b}_i-1)\,\, \mathbf{x}(\tilde{b}_i)\,\,\mathbf{x}(\tilde{b}_i+1) \dots \mathbf{x}(\tilde{b}_i+R_m) \big]^t,
\end{align}
where the superscript $t$ means taking the transpose. Note that the $i$-th R peak is located on the $(L_m+1)$-th entry of all $\mathbf{x}_i$. Denote $\mathcal{N}_m^{(i)}=\{\mathbf{x}^{(i_1)},\ldots,\mathbf{x}^{(i_{K_m})}\}$ to be the first $K_m$ nearest neighbors of $\mathbf{x}^{(i)}$ with respect to the $L^2$ norm, where $K_m\in\NN$ is chosen by the user. The $i$-th maternal cardiac activity is thus estimated by
\begin{align}
\tilde{\mathbf{x}}_{m}^{(i)}(l):=\text{median}\{\mathbf{x}^{(i_j)}(l)\}_{j=1}^{K_m},
\end{align}
where $l=1,\ldots,L_m+R_m+1$.

Before estimating the maECG from $\{\tilde{\mathbf{x}}_{m}^{(i)}\}_{i=1}^{M_m}$, we need to take care of the possible overlapping issue. If $\tilde{\mathbf{x}}_{m}^{(i)}$ overlap with its neighboring segments $\tilde{\mathbf{x}}_{m}^{(i+1)}$, we need to taper the overlapping regions of both segments. See Algorithm \ref{alg:nonlocalmedian} for the implementation of the tapering step. To simplify the notation, we use the same notation $\tilde{\mathbf{x}}_{m}^{(i)}$ to denote the tapered segment.
Finally, the maECG could be estimated by $\tilde{\mathbf{x}}_{m}\in\RR^N$, where
\begin{equation}
%\tilde{\mathbf{x}}_{m}(\hat{B}(i)+j)=\tilde{\mathbf{x}}_{m}^{(i)}(m+1+j)
\tilde{\mathbf{x}}_{m}(\hat{b}_i+j) = \sum^{M_m}_{i=1}\tilde{\mathbf{x}}_{m}^{(i)}(j+L_m+1)
\end{equation}
for all $i=1,\ldots,M_m$ and $j=-L_m,\ldots,R_m$, and zero otherwise. In practice, we found that the result is stable when $K_m$ ranges from 20 to 60. The whole procedure is sketched in Algorithm \ref{alg:nonlocalmedian}.

\begin{algorithm}[h!]
\begin{algorithmic}
\STATE[INPUT] the aECG signal $\mathbf{x}\in \RR^N$; an estimated R-peak sequence $\tilde{B}_m$; $L_m$ and $R_m$; the number of nearest neighbors $K_m$;
\STATE[STEP 3-0] Initialize $\mathbf{y}\in \RR^N$, $\mathbf{y}(i)=0$ where $i=1,\dots,N$
\STATE[STEP 3-1] Segmenting the aECG signal according to the R peaks
\FOR {$i = 1$ \TO $M_m$}
\STATE $I_{m,i} := \{\tilde{b}_i-L_m, \tilde{b}_i-L_m+1, \dots, \tilde{b}_i+R_m\}$
\STATE $\mathbf{x}^{(i)} := \big[ \mathbf{x}(\tilde{b}_i-L_m) \dots \mathbf{x}(\tilde{b}_i-1)\,\, \mathbf{x}(\tilde{b}_i)\,\,\mathbf{x}(\tilde{b}_i+1) \dots \mathbf{x}(\tilde{b}_i+R_m) \big]^t$;
\ENDFOR
\STATE $\mathbf{X} := \{\mathbf{x}^{(i)}\}^{M_m}_{i=1}\subset \RR^{L_m+R_m+1}$
\FOR {$i = 1$ \TO $M_m$}
%\STATE[STEP2] Compute $d(\mathbf{x}^{(i)}, \mathbf{x}^{(j)})$, the Euclidean distance matrix
%\FOR {$j=1$ \TO $M_m$}
%\STATE $d(\mathbf{x}^{(i)}, \mathbf{x}^{(j)})=\sqrt{\sum^{L}_{i=1}|\mathbf{x}^{(i)}-\mathbf{x}^{(j)}|^2}$
%\ENDFOR
\STATE[STEP 3-2] Select $\mathbf{X}_{K_m}\subseteq \mathbf{X}$, the $K_m$-th nearest neighbors of $\mathbf{x}^{(i)}$ in the $L^2$ distance.
\STATE[STEP 3-3] Take non-local median: $\tilde{\mathbf{x}}_{m}^{(i)}(l)=\mathrm{median}\{\mathbf{x}^{(k)}(l)\}_{\mathbf{x}^{(k)}\in\mathbf{X}_{K_m}}$, where $l=1,\ldots,L_m+R_m+1$.
\STATE[STEP 3-4] Taper the (possible) overlap region between $\tilde{\mathbf{x}}_{m}^{(i)}$ and $\tilde{\mathbf{x}}_{m}^{(i-1)}$, and between $\tilde{\mathbf{x}}_{m}^{(i)}$ and $\tilde{\mathbf{x}}_{m}^{(i+1)}$
\STATE $o_1:=|I_{m,i-1}\cap I_{m,i}|$, $o_2:=|I_{m,i}\cap I_{m,i+1}|$
\FOR {$l=1$ \TO $L_m+R_m+1$}
\IF {$l\leq o_1$}
\STATE $\tilde{\mathbf{x}}_{m}^{(i)}(l)\leftarrow\tilde{\mathbf{x}}_{m}^{(i)}(l)\sin^2(\frac{\pi l}{2o_i})$
\ENDIF
\IF {$l\geq (L_m+R_m+1)-o_2+1$}
\STATE $\tilde{\mathbf{x}}_{m}^{(i)}(l)\leftarrow\tilde{\mathbf{x}}_{m}^{(i)}(l)\cos^2(\frac{\pi (l+o_2-1-(L_m+R_m+1))}{2o_i})$
\ENDIF
\ENDFOR
\STATE[STEP 3-5] $\mathbf{y}|_{I_{m,i}}\leftarrow\mathbf{y}|_{I_{m,i}}+\tilde{\mathbf{x}}_{m}^{(i)}$
\ENDFOR
\STATE[OUTPUT] the maECG signal $\mathbf{y}\in\RR^N$.
\end{algorithmic}
\caption{Non-local median for maECG extraction.}
\label{alg:nonlocalmedian}
\end{algorithm}

\subsubsection{Final step: get the fIHR and obtain the fECG signal}

Denote $\tilde{\mathbf{x}}_{f,0}:=\mathbf{x}-\tilde{\mathbf{x}}_{m}$ to be the {\em rough fECG} estimate. %Note that with the optional step 1 in the detection stage, we also have an estimate of fIHR, but we .
%However, in practice, due to the possible noise and other artifacts,
The fIHR, the fetal R peaks, and fECG could be obtained by repeating Step 1, Step 2, and Step 3 on the rough fECG estimate. Precisely, the fIHR could be obtained by running the de-shape STFT on $\tilde{\mathbf{x}}_{f,0}$ and another DP curve extraction on $\mathbf{W}_{\tilde{\mathbf{x}}_{f,0}}$ in Step 1.

[Optional step 2] Before running the DP curve extraction, we could re-weight $\mathbf{W}_{\tilde{\mathbf{x}}_{f,0}}$ around the band associated with the mIHR $\eta_m$ by a small constant $\theta_f>0$ to reduce to the possible impact of the remaining maECG component. Precisely, set $\mathbf{W}_{\tilde{\mathbf{x}}_{f,0}}(n,j_n)$ to be $\theta_f\mathbf{W}_{\tilde{\mathbf{x}}_{f,0}}(n,j_n)$, where $n=1,\ldots,N$ and $j_n\in\{\eta_m(n)-N_\Delta,\eta_m(n)-N_\Delta+1,\ldots,\eta_m(n)+N_\Delta-1,\eta_m(n)+N_\Delta\}$. In practice, choosing $\theta_f=1/10$ could slightly improve the result, and we recommend to take it into account.

Denote $\eta_f\in\RR^N$ to be the final  estimated fIHR. The fetal R peak location could be further refined by running the beat tracking technique by the DP. Denote the final estimated fetal R peaks by a strictly increasing sequence $\tilde{B}_f$. The fECG signal could be reconstructed by the nonlocal median algorithm, with $L_f,R_f\in\NN$ chosen based on the physiological fact and $K_f\in\NN$ nearest neighbors chosen by the user. Denote the final reconstructed fECG as $\tilde{\mathbf{x}}_{f}\in\RR^N$. In practice, we found that the result is stable when $K_f$ ranges from 20 to 60.

[Optional step 3] This optional step takes the physiological constraint into account. If the average $\eta_f$ determined from the rough fECG signal is smaller than $\eta_m$, the final reconstructed maECG and fECG could be exchanged to respect the physiological constraint that normally the mIHR is slower then fIHR. When both the fetus and mother are healthy, this step could help when the fECG is strong than the maECG, a case which is not commonly seen.

\subsection{Evaluation statistics}

The R peak detection result is evaluated by beat-to-beat comparisons between the detected beats and the annotation provided by the experts. We follow the suggestion \cite{Guerrero-Martinez2006} and choose a matching window of 50 ms. Denote $TP$, $FP$, $TN$, and $FN$ to be true positive, false positive, true negative, and false negative. 
We report the sensitivity (SE)
\[
SE=TP/(TP+FN),
\]
the positive predictive value (PPV)
\[
PPV=TP/(TP+FP),
\]
and the $F_1$ score, which is the harmonic mean of PPV and SE,
\[
F_1=2TP/(2TP+FN+FP).
\]
In addition, the mean absolute error (MAE) of estimated R peak locations is also report. To make the evaluation independent of the detection accuracy, we follow the suggestion in \cite{Andreotti2016} to calculate the MAE only on TP annotations.

To evaluate the fECG morphology recovery, we consider the correlation between the estimated fECG signal and the true fECG signal, over the period ranging from 80 msec before the annotated R peak and 120 msec after the annotated R peak.  Again, we evaluate the correlation only on TP annotations.

\section{Material and Result}\label{Section:MaterialResult}

For a fair comparison, the parameters for the algorithm are set to be the same for all signals throughout the paper, and the optional step 2 and 3 are implemented, unless otherwise stated. The window $h$ is chosen to be the Hamming window of length $5$ second, the up-sampling rate $\alpha=10$, the frequency resolution in STFT and de-shape STFT is set to $0.02$ Hz, the quefrency resolution is set to $10$ msec, $\gamma=0.3$ for the STCT, and $\upsilon=10^{-4}$\% of the root mean square energy of the signal under analysis for the de-shape STFT. In the beat tracking, $\lambda_{\texttt{BT}}$ is set to $50$.
In the nonlocal median, we choose $K_m=K_f=40$.

\subsection{Material}

We evaluate the proposed algorithm on three databases of aECG signals.
The first database is the simulated fECG signal database ({\em fecgsym}) \cite{Andreotti2016}. The publicly available simulator generates simultaneously the maECG and fECG signals in 34 channels, which model a number of realistic non-stationary physiological phenomena that affect the morphology and dynamics of the aECG, including different kinds of noise. A total of seven physiological events are introduced in the simulator:
\begin{enumerate}
\item Baseline: the baseline abdominal mixture with a constant fIHR and mIHR without noise or events;
\item Case 0: baseline signal contaminated by noise;
\item Case 1: Case 0 is complicated by the fetal movement noise;
\item Case 2: the fIHR and mIHR are time-varying with noise;
\item Case 3: Case 2 is complicated by the uterine contraction noise;
\item Case 4: Case 2 is complicated by ectopic beats in both fECG and maECG;
\item Case 5: twin pregnancy contaminated by noise.
\end{enumerate}
The simulator also generates five different levels of additive noise, ranging from 0, 3, 6, 9, to 12 dB. Each physiological event and noise level were simulated independently five rounds to mimic the realistic situation.
A generated benchmark is available in \url{https://physionet.org/physiobank/database/fecgsyndb/} \cite{Goldberger_Amaral_Glass_Hausdorff_Ivanov_Mark_Mietus_Moody_Peng_Stanley:2000}, which contains ten subjects, five rounds, and five SNR's for each case. Each simulation is of 5-minute long and is discretized at the sampling rate 250Hz.
For each subject, case, round, and SNR, there are one maECG, fECG, two noise realizations, and one extra uterine contraction noise in Case 3 in the database. We sum all these time series together to get the simulated signal for analysis.
We test our algorithm on all Cases, all levels of noise, and all five simulations, except Case 5. We consider the twin pregnancy case as an independent project, and the result will be reported in the other work.
The second database is the PhysioNet non-invasive fECG database ({\em adfecgdb}), where the aECG signals with the annotation provided by experts is publicly available \url{https://www.physionet.org/physiobank/database/adfecgdb/}\cite{Goldberger_Amaral_Glass_Hausdorff_Ivanov_Mark_Mietus_Moody_Peng_Stanley:2000}.
The annotation is determined from the direct fECG recorded from the fetal scalp lead.
There are five pregnant women between 38 to 40 weeks of pregnancy in this database, each has 4 aECG channels and one direct fECG signal. 
The signal is of 5-minute long, and is digitalized at 1000Hz with 16bit resolution. More details could be found in \url{https://www.physionet.org/physiobank/database/adfecgdb/}.

The third database is the 2013 PhysioNet/Computing in Cardiology Challenge (\url{https://physionet.org/challenge/2013/#data-sets}), abbreviated as CinC2013, which is composed of three sets, learning (training) set A, open test set B, and hidden test set C, where only set A comes with the annotation. We thus used set A for an assessment of our proposed algorithm. There are 75 subjects in set A, and for each subject four aECG channels of length 1-minute long sampled at 1000 Hz are provided. More details about this database could be found in \url{https://physionet.org/challenge/2013/#data-sets}. 

\subsection{Results of the simulation: fecgsym database}

In this simulated database, since we a priori know that the noise could be as big as 0 dB, so we choose a longer window of length 10 s. The other parameters are fixed for all the cases.

Following the recommended report method in \cite{Andreotti2016}, Tables \ref{tab: fecgsynF1} and \ref{tab: fecgsynF1part2} give an overview of the performance of the proposed algorithm. In addition to reporting the 1min result, we also report the whole 5min result.
For the 1min (respectively 5min) result, for each subject, noise level, case, and round, we find the channel and the 1-minute subset out of the 5-minute signal (respectively the channel and the whole 5-minute signal) that gives the highest $F_1$. For each noise level and case, the median and interquartile (IQR) of those highest $F_1$ values, and the associated MAE of all subjects and rounds are shown in Tables \ref{tab: fecgsynF1}. 
To better understand the performance of the proposed algorithm, the whole gross statistics with the \textit{tenth} highest $F_1$ is reported in Table \ref{tab: fecgsynF1part2}.

From these two tables, we could see that as SNR decreases, the performance decreases,
which is consistent with the results of different TS techniques reported in \cite[Figure 4]{Andreotti2016}. It is also clear that the overall accuracy of the proposed algorithm is higher than those of TS techniques and adaptive filtering techniques reported in \cite[Table 3]{Andreotti2016}. 
We also notice that the IQR in our report is higher than those of different TS techniques reported in \cite[Table 3]{Andreotti2016}. This is caused by the fact that we do not use the ground truth maternal R peak information but estimate it from the aECG signal.\footnote{Note that the purpose of \cite{Andreotti2016} is comparing different methods, ranging from different BSS algorithms to single-lead TS algorithms, instead of evaluating simply a specific TS algorithm, but in this paper we only focus on evaluating our proposed algorithm.}
We should notice that when SNR decreases, the lost information could not be recovered by only one channel signal.
Clearly, in Case 3 where the signal is contaminated by the uterine contraction noise, the results on the 5-minute long signal is much lower. This comes from the fact that the energy of the uterine contraction is much larger than both the maECG and fECG, and almost no information could be extracted when the uterine contraction happens.
Note that except Case 4, the MAE outperforms the reported results in \cite[Table 3]{Andreotti2016}. This indicates the strength of beat tracking and nonlocal median. In Case 4, although the MAE is larger than the other cases, it is still on the same level of that reported in \cite[Table 3]{Andreotti2016}.

For the computational time over the 5-minute signal, it takes about 12 seconds to finish a round in MacBook Pro (Retina, 15-inch, Mid 2014) with Processor 2.5GHz Intel Core i7, Memory 16 GB 1600MHz DDR3, OS X Yosemite (Version 10.10.5), and Matlab R2014b without implementing the parallel computation.

\begin{figure}
\centering
\includegraphics[width=0.98\columnwidth]{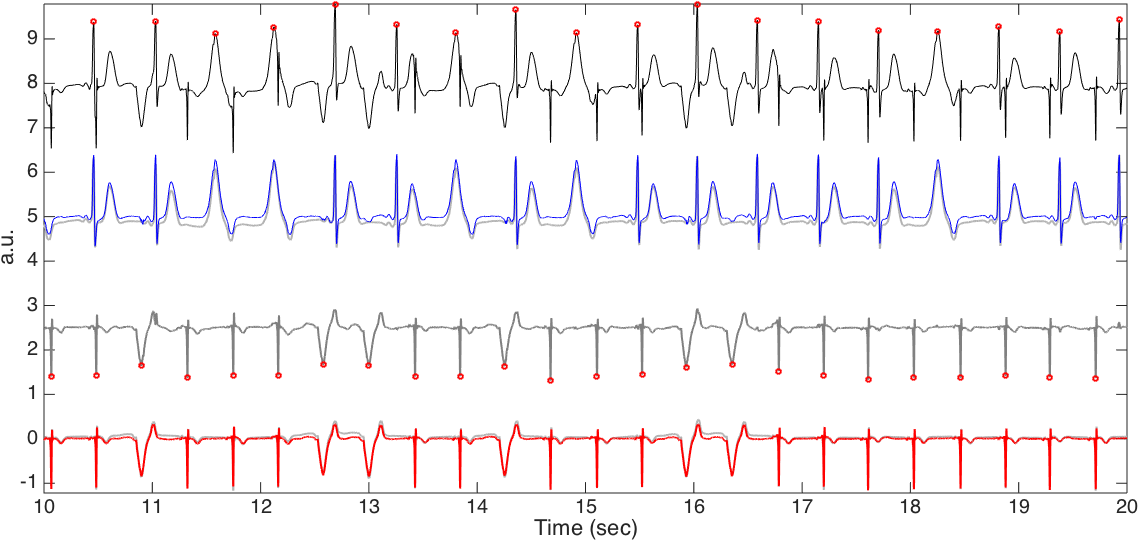}\\
\includegraphics[width=0.98\columnwidth]{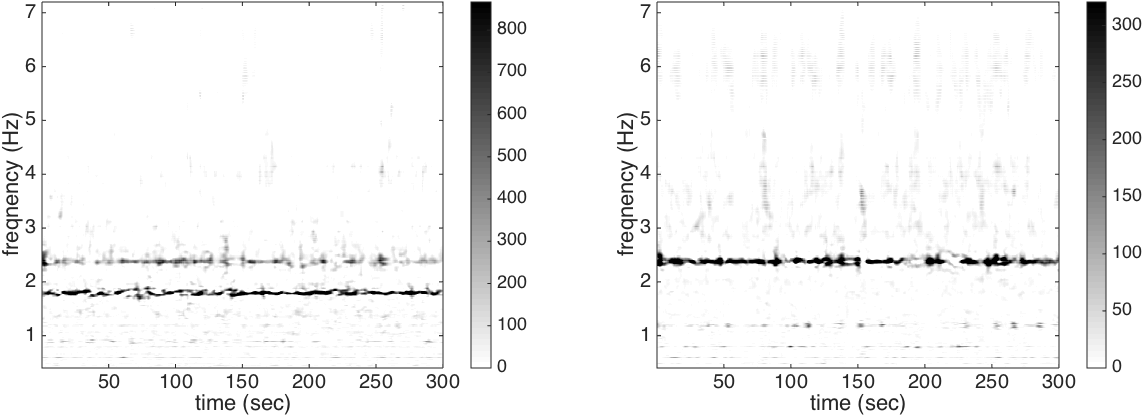}
\includegraphics[width=0.98\columnwidth]{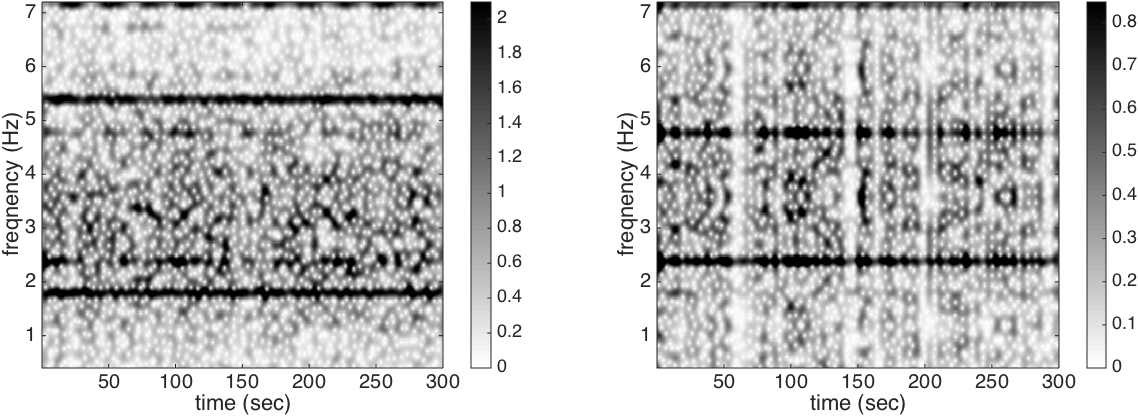}
\caption{The result of subject 1, round 1, case 4, and channel 21 without noise. In the top subplot, all relevant signals are shown together for a visual comparison. The clean aECG signal is shown in black  (shifted up by 8 units)  with the detected R peaks in round red circles on the top row; the clean maECG is shown in light gray  (shifted up by 5 units)  superimposed with the estimated maECG signal in blue on the second row; the rough fetal ECG is shown in dark gray  (shifted up by 2.5 units)  with the detected R peaks in round red circles on the third row;  the clean fECG is shown in light gray superimposed with the estimated fECG signal in red on the bottom row. In the middle subplot, the de-shape STFT of the maECG is shown on the left and the de-shape STFT of the rough fECG is shown on the right. In the bottom subplot, the STFT of the maECG is shown on the left and the STFT of the rough fECG is shown on the right. Note that the discrepancy between the clean maECG and the estimated maECG comes from the median filter, which in general is needed to remove the baseline wandering.}
\label{fig:simu_dsSTFT_clean}
\end{figure}

The fECG morphology estimation results are shown in Table \ref{tab: fecgsynMorphology} and Figure \ref{fig:simu_dsSTFT_clean}. The high correlation between the estimated fECG and the ground truth indicates that the nonlocal median algorithm leads to an estimated fECG morphology with low distortion. Note that in this evaluation, we only evaluate the correlation on the correctly detected fECG beats (true positive fECG beats). This suggests that if we could accurately estimate the fetal R peaks (the higher $F_1$ score), then we could have an accurate fECG reconstruction.

To look deeper into the algorithm and its performance, we take subject 1, round 1, case 4, and channel 21 into account, and show the result without noise in Figure \ref{fig:simu_dsSTFT_clean}. In this example, over the 5-minute, the $F_1$ is 1 and the MAE is 0.78 msec.
In addition to the de-shape STFT of the maECG signal, the estimated R peaks by the beat tracking, the decomposed maternal ECG, the rough fECG estimate $\mathbf{x}_{f,0}$ and its TFR, and the final fECG estimate are shown.
We see that by the de-shape STFT, the information of non-sinusoidal oscillation; that is, the time-varying wave-shape function, and the IHR information are decoupled, and only the IHR information is shown in the TF representation. Note that we could see both the mIHR and fIHR in the TFR, and as expected, fIHR has a weaker intensity than mIHR, meaning that the energy of fECG is smaller. The intensity level could be seen in the colorbar. For a comparison, we could see that in the STFT of aECG, both the mIHR and fIHR could be seen, while the fIHR is much weaker compared with that in the de-shape STFT. Furthermore, the multiples of the maternal fundamental component could mask both the mIHR and fIHR information and even interfere with each other. 
To show how the fECG could be reconstructed, the estimated fECG and the ground truth fECG signal are put side by side for a comparison. 
On the other hand, we could see that the nonlocal nature of the nonlocal median algorithm does help us to recover ECG morphology, even the ectopic beats, and the median and IQR of the correlation of all TP beats are $0.997$ and $0.004$.

\begin{figure}
\centering
\includegraphics[width=0.98\columnwidth]{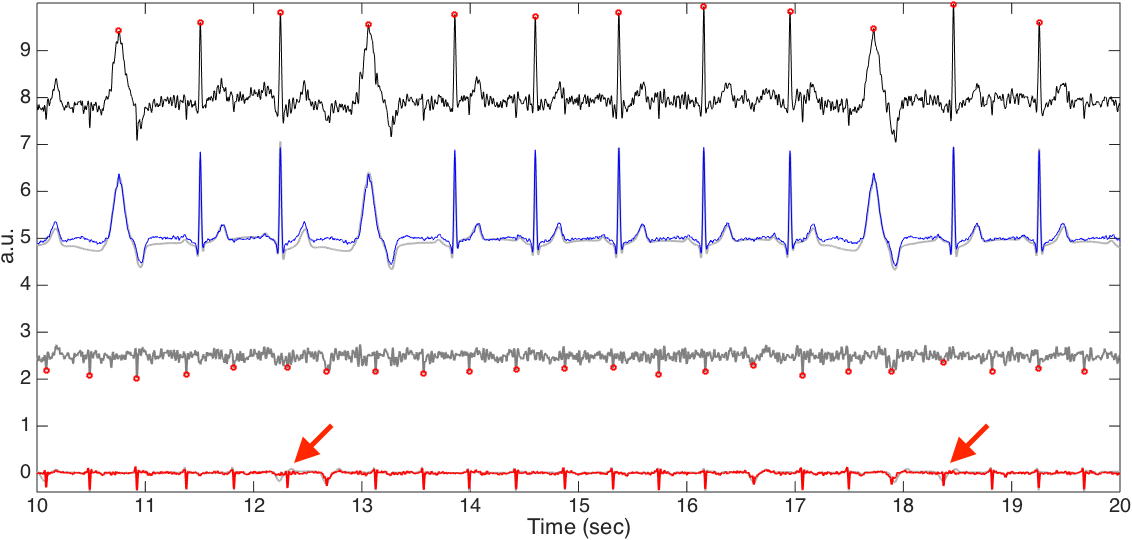}\\
\includegraphics[width=0.98\columnwidth]{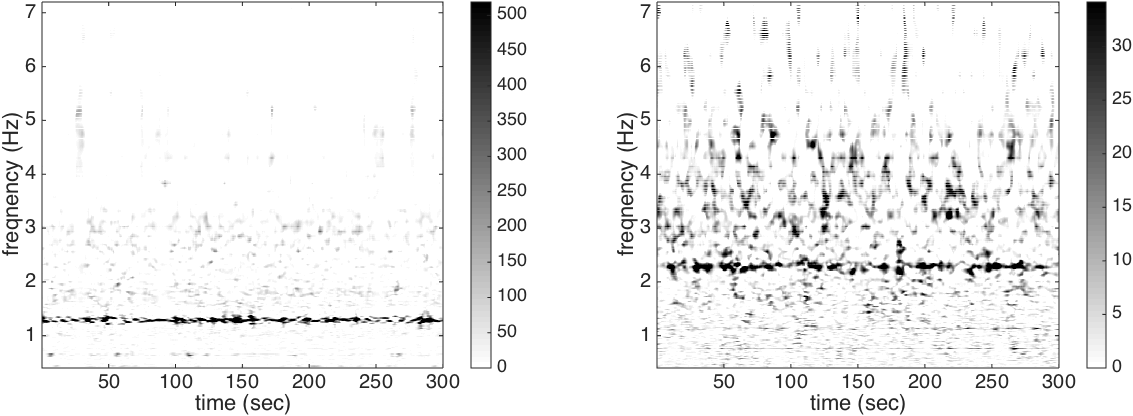}
\includegraphics[width=0.98\columnwidth]{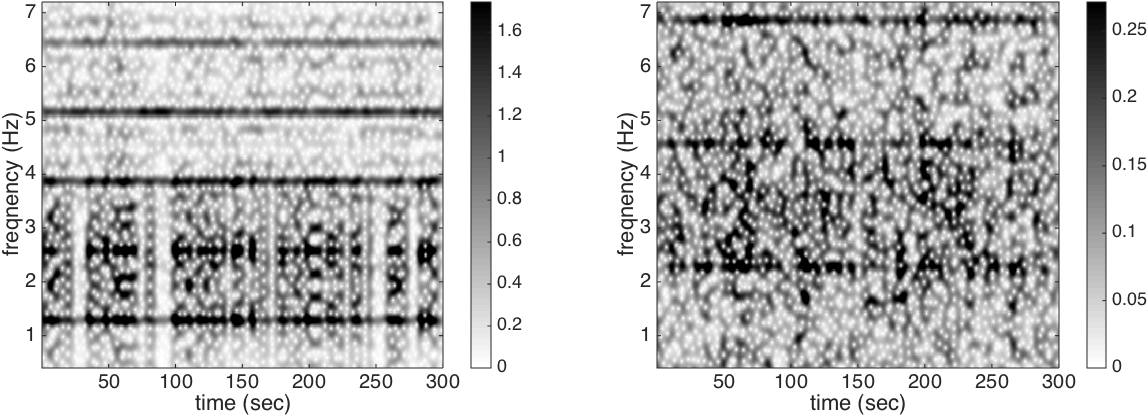}
\caption{The result of subject 3, round 1, case 4, and channel 23 with the SNR 6dB. In the top subplot, all relevant signals are shown together for a visual comparison. The clean maECG signal is shown in black (shifted up by 8 units) with the detected R peaks in round red circles on the top row; the clean maECG is shown in light gray superimposed with the estimated maECG signal in blue  (shifted up by 5 units)  on the second row; the rough fetal ECG is shown in dark gray  (shifted up by 2.5 units)  with the detected R peaks in round red circles on the third row;  the clean fECG is shown in light gray superimposed with the estimated fECG signal in red on the bottom row.
The two red arrows indicate two ectopic beats that are not recovered by the nonlocal median algorithm. 
In the middle subplot, the de-shape STFT of the maECG is shown on the left and the de-shape STFT of the rough fECG is shown on the right.
In the bottom subplot, the STFT of the maECG is shown on the left and the STFT of the rough fECG is shown on the right.  Note that the discrepancy between the clean maECG and the estimated maECG comes from the median filter, which in general is needed to remove the baseline wandering.}
\label{fig:simu_dsSTFT_9dB}
\end{figure}

To show the result when noise exists, we take another signal, subject 3, round 1, case 4, and channel 23 into account, and with SNR 6dB, as an example, and show the result in Figure \ref{fig:simu_dsSTFT_9dB}. We choose this signal as an example since it has a smaller fECG amplitude. As can be clearly seen, even when the noise is 6dB, the de-shape STFT still gives a reasonable TFR with the IHR information, although there are several ``speckles'' in the background, which come from the noise. In this example, since fECG is smaller compared with the maECG, the fIHR could not be clearly seen in the de-shape STFT of the aECG. However, it could be seen clearly in the de-shape STFT of the rough fECG estimate. The intensity level could be seen in the colorbar. Also note that since the fECG is smaller and the noise is big, and no denoise technique is combined into the algorithm, the de-shape STFT of the rough fECG estimate has a ``scattered'' background. For the morphology reconstruction, due to the large noise, the nonlocal median failed to recover two ectopic beats, as are indicated in the red arrows. Note that although the nonlocal nature of the nonlocal median has the power to handle ectopic beats, it is not designed for this purpose. To have a better recovery of the ectopic beats, more features specifically designed for ectopic beats should be taken into account. In this case, the $F_1$ is 0.996, the MAE is 3.476 msec,
and the median and IQR of correlations of all TP beats are $0.968$ and $0.034$. We could see that the nonlocal median provides a convincing potential in recovering the fECG morphology, specially when noise exists.

\subsection{Results of the first real database: adfecgdb database}

Table \ref{tab: adfecgdbF1} shows the $F_1$, MAE, PPV and SE of all channels and all subjects. Notice that in the r10 record, the direct fECG measurement was lost between 187 and 191 s, and between 203 and 211 s, so these two segments were neglected in the evaluation. To avoid the boundary effect introduced by the window function, the first and last 0.5 second in every recording are not evaluated neither. 
In terms of $F_1$, compared with the state-of-art result reported in, for example, \cite{CastilloMorales2013}, our result is overall better. The MAE, which is less reported in the literature, is as small as $10$ msec, which indicates the potential to carry out the fetal HRV analysis from the single-lead aECG, although this topic is out of the scope of this paper. However, as compared with the simulated database, the MAE is larger. Note that this larger MAE is reasonable, since in this database, we use the R peaks of the direct fECG as our ground truth, but the direct fECG has a different projection direction compared with the fECG recorded from the maternal abdomen. This difference leads to the slightly larger MAE.

To further explore the proposed algorithm, in Figure \ref{fig:real_deSTFT}, we show the result of  STFT and de-shape STFT of the third channel of the case r07, denoted as $\mathbf{x}_{11}$, as well as the the result of STFT and de-shape STFT of the rough fECG estimate from $\mathbf{x}_{11}$, denoted as $\mathbf{x}_{11,f0}$ (the signal used in the flowchart in Figure \ref{FlowChart} is the channel 1 of the case r01). Clearly, in the STFT of $\mathbf{x}_{11}$, we not only see the fundamental IF of maECG around $1.2$ Hz, but also its multiples, due to the nature of non-sinusoidal oscillation of the ECG signal. We could also see a relatively vague component around 2 Hz in STFT, which turns out to be the fIHR. On the other hand, after the de-shape process, in the de-shape STFT of $\mathbf{x}_{11}$, only two dominant curves associated with the mIHR and fIHR are left. 

An example of the estimated fECG waveform, $\mathbf{x}_{11,f0}$, is shown in Figure \ref{fig:real_morphology}. The R peaks are clearly well reconstructed and match those in the direct fECG signal recorded from the fetal scalp. By comparing the rough fECG estimate and the final fECG estimate, we could see the effectiveness of nonlocal median. Furthermore, some fiducial points could be observed, as is shown in Figure \ref{fig:real_morphologyZoom}, which is the zoomed in of Figure \ref{fig:real_morphology}. Although not all critical fiducial points could be extracted, this result indicates the potential of studying the morphology of fECG, particularly taking into account the fact that we only count on a single-lead aECG signal. We mention that in some cases the P wave and T wave could be reconstructed, but in general they are buried in the noise, so we do not consider it as an achievement of the proposed algorithm, and more work is needed to recover these morphological features.

\subsection{Results of the second real database: CinC2013 database}

In this database, since the signal is of only 1-minute long, we take $K_m=K_f=10$ to enhance the local nature of the nonlocal median algorithm. In practice, the result is slightly worse if $K_m=K_f=40$.

Here we follow the report suggested in \cite{BeharOsterClifford2014} to report the mean in addition to the median of all subjects. For each subject, we choose the channel with the highest $F_1$ score as the channel for this subject, and report the gross statistics of the obtained $F_1$, PPV, SE, and MAE over 75 subjects. Over 75 subjects, the mean and standard deviation of the $F_1$ score (respectively PPV and SE) are $86.37$ and $22.9\%$ (respectively 85.77\% and 23.42\%, and 87.5\% and 22.05\%), and the mean and standard deviation of MAE are $6.12$ msec and $5.56$ msec.
\footnote{If we follow \cite{BeharOsterClifford2014} and remove 7 cases, including a33, a38, a47, a52, a54, a71, and a74, then the mean and standard deviation of $F_1$ score (respectively PPV and SE) are $88.73\%$ and $20.96\%$ (respectively $88.39\%$ and $21.08\%$, and $89.12\%$ and $20.77\%$), and the mean and standard deviation of MAE over 75 subjects are $5.88$ msec and $5.14$ msec.} Our $F_1$ result is compatible with the reported results in \cite{Ghaffari2015,LeeLee2016,DiMariaLiu2014,BeharOsterClifford2014,AndreottiRiedl2014}, while the MAE is better.
On the other hand, over 75 subjects, the median and IQR of the $F_1$ score (respectively PPV and SE) are 98.28\% and 14\% (respectively 98.25\% and 14.72\%, and 98.59\% and 22.05\%), and the median and IQR of MAE are $3.81$ msec and $5.74$ msec.

The discrepancy between the mean and median indicates the existence of outliers in the database. We thus take a closer look into the database. We found that the fECG signal could be hardly visualized in 10 subjects in the database, including a27, a32, a43, a50, a59, a60, a63, a64, a68, and a75, even when the signal is clean, and these cases are considered difficult for our algorithm. If we remove these cases, the
mean and standard deviation of the $F_1$ score (respectively PPV and SE) become $94.54\%$ and $11.63\%$ (respectively 93.87\% and 13.05\%, and 95.78\% and 8.79\%), and the mean and standard deviation of MAE become $4.35$ msec and $3.04$ msec.
On the other hand, the median and IQR of the $F_1$ score (respectively PPV and SE) become $98.84\%$ and $4.91\%$ (respectively 98.53\% and 6\%, and 99.21\% and 4.16\%), and the mean and standard deviation of MAE become $3.35$ msec and $4.33$ msec.

To have a deeper look into the difficulty, we show one example, a59, which is considered difficult case for our algorithm, in Figure \ref{fig:CinC2013a59part1}. In Figure \ref{fig:CinC2013a59part1}, the second channel of a59 is shown for an illustration. It is clear that the signal is quite clean, and it is not easy to identify if the fECG exists, even with the help of the provided annotation. Not surprisingly, the fetal R peaks are all detected incorrectly. By a direct visual inspection, we could see that those locations that are erratically identified as fetal R peaks are nothing but the residue coming from the incomplete maECG removal. This fact could also be observed in the de-shape STFT of the rough fECG estimate -- the only ``dominant component'' in the de-shape STFT coincides with that of the aECG, as is indicated by the red arrow. We mention that the other three channels all share the same result -- the fECG is too small to be even sensed from this relatively clean aECG signal.
However, if we take the VCG notion into account, then it is possible to enhance the result. Precisely, by linearly combining different channels, we have a chance to obtain a stronger fECG signal relative to the maECG signal, and hence the result could be better. This idea is shown in \ref{fig:CinC2013a59part2}, where the difference between channel 2 and channel 3 is analyzed. Clearly, we could see that now the fetal R peaks could be almost perfectly recovered. In this specific case, the $F_1$ is $99.3\%$ and the MAE is $0.993$ msec. We mention that if we do the same linear combination trick between channel 1 and channel 3 (respectively channel 2 and channel 3, channel 2 and channel 3, channel 1 and channel 2, channel 1 and channel 2, channel 2 and channel 3, channel 2 and channel 3, channel 2 and channel 3, and channel 1 and channel 2), the $F_1$'s of a27 (respectively a32, a43, a50, a60, a63, a64, a68, and a75) become 79.3\% (respectively 100\%, 100\%, 97.9\%, 84.7\%, 91\%, 98.5\%, 91.4\%, and 86.2\%). While this is a naive way to handle the problem and it works well in this preliminary analysis, however, we need at least two channels. Since we focus on the single-lead aECG analysis in this paper, this direction will be explored in the other work. We mention that this idea is also considered in \cite{AndreottiRiedl2014}.

% DiMariaLiu2014 use 100ms matching window and remove 9 cases

\section{Discussion and future work}\label{Section:Discussion}

In this section, we discuss the difficulty of the fECG extraction problem from the signal processing viewpoint, different findings of the proposed algorithm, and several possible applications and future works.

\subsection{General technical difficulty}\label{Subsection:TechnicalDifficulity}

In the analysis of multi-component oscillatory signals, there are two chicken-and-egg problems of fundamental importance. The first is the {\em detection problem}; that is, how to determine the number of components and how to find the fundamental frequency or equivalently the fundamental period of each component? The second one is the {\em separation problem}; that is, how to separate all components from a recorded signal? These two problems coexist in many kinds of data, ranging from physiological signals to polyphonic music signals, where each component in the mixture contains information different from others.

Previously, these two problems are usually discussed separately, probably because only discussing one of them is challenging enough. However, recently some research works start to consider these two problems as a single one by viewing the separation problem as an extension of the detection problem. For example, we could simultaneously estimate the IF of each component, and then extract the wave-shape function as well as each component. 

The methods could be classified into two classes, {\em iterative} or {\em joint}. The iterative approach extracts the most prominent IF/component in each iteration, until no additional IF/component can be found. Although iterative models are usually computationally inexpensive, they have a main drawback: iterative models tend to accumulate errors at each iteration step if the feature representation is not robust enough. 
To handle this limitation, we could consider the ``joint'' approach. 
Vast majority of recent approaches in multi-pitch estimation and source separation in music now falls within the ``joint'' category \cite{benetos2013automatic}, and more and more studies on source separation started to consider pitch information to improve the source separation algorithms \cite{chan2015vocal, ikemiya2014mirex}.
Note that although it is intuitive to utilize the IF information to handle the detection and separation problem, this kind of approach has been less studied until recent years, probably because the task of finding IF is by no means easy, especially when there are multiple components. Compared with the iterative method, the joint methods lead to more accurate estimates but with more involved mathematical tool and increased computational cost.

\begin{figure}
\centering
\includegraphics[width=0.49\columnwidth]{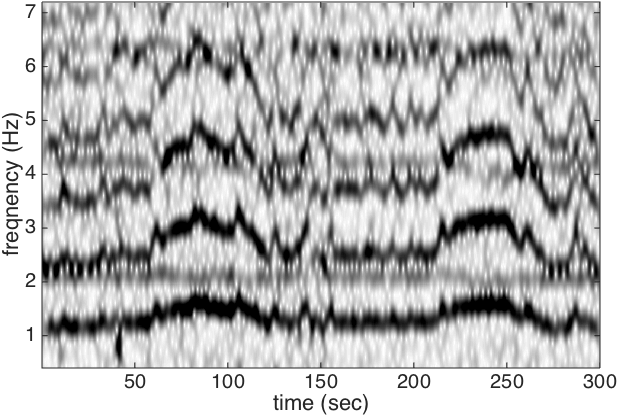}
\includegraphics[width=0.49\columnwidth]{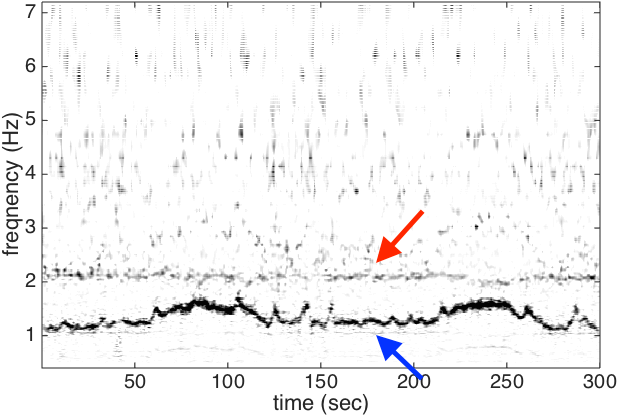}\\
\includegraphics[width=0.49\columnwidth]{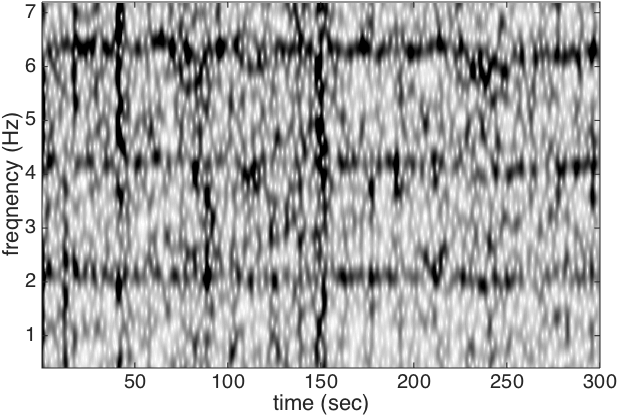}
\includegraphics[width=0.49\columnwidth]{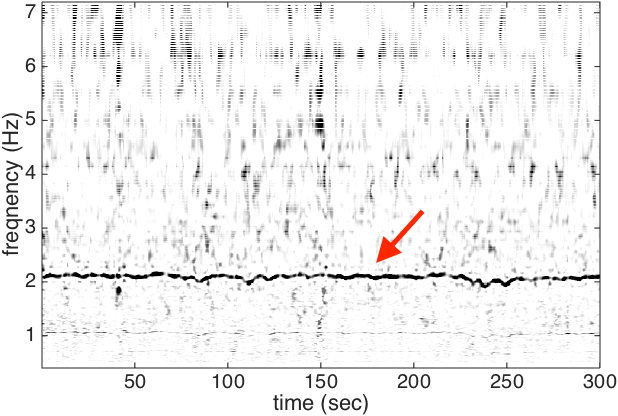}
\caption{Upper left and upper right: the STFT and de-shape STFT of the first channel of the case r01; bottom left and bottom right: the STFT and de-shape STFT of the rough fECG estimate. The red arrow indicates the fIHR, and the blue arrow indicates the mIHR.}
\label{fig:real_deSTFT}
\end{figure}

\begin{figure}
\centering
\includegraphics[width=0.98\columnwidth]{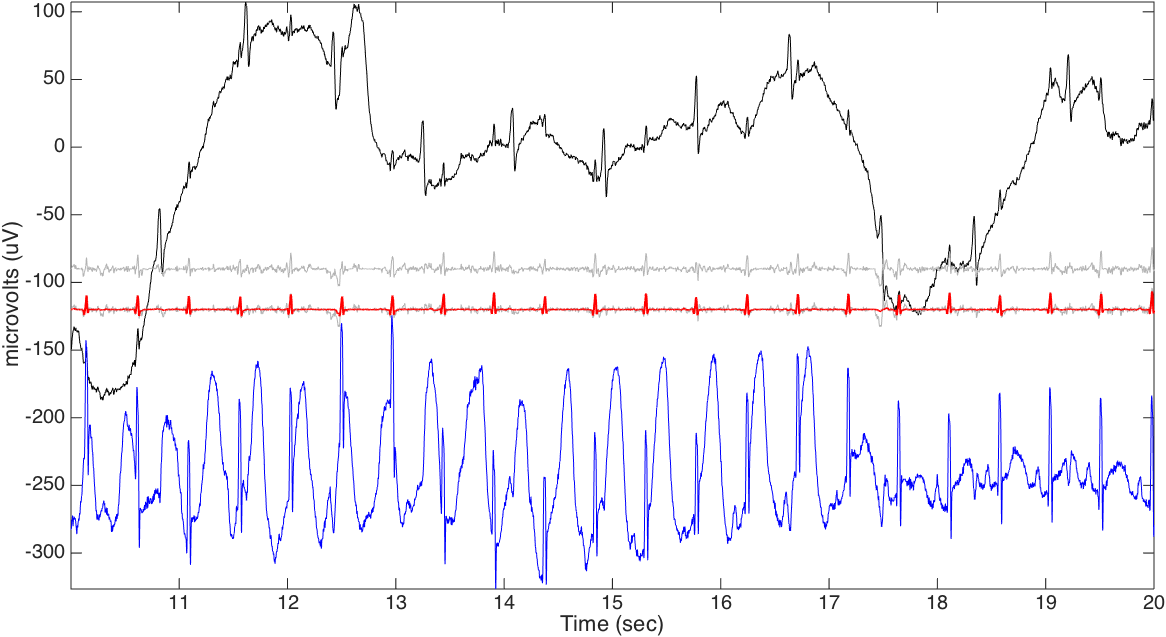}
\caption{The aECG signal of the third channel of the case r07 is plotted in black on the top row. The rough fECG signal estimate (shifted down by 90 units) is plotted in gray on the second row. The final fECG estimation (shifted down by 120 units) is plotted in red superimposed on the rough fECG signal (shifted down by 120 units) plotted in gray for the comparison purpose on the third row. The direct fECG recorded from the fetal scalp (shifted down by 240 units) is shown in blue on the bottom row.}
\label{fig:real_morphology}
\end{figure}

\begin{figure}
\centering
\includegraphics[width=0.7\columnwidth]{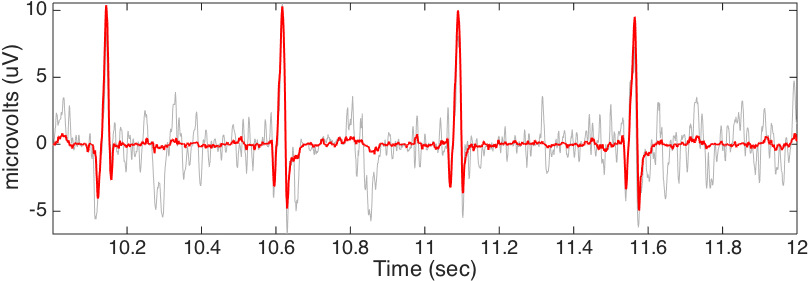}
\caption{The small segment fECG estimation is plotted in red with the rough fECG signal plotted in gray for the comparison purpose. The data is the third channel of the case r07.}
\label{fig:real_morphologyZoom}
\end{figure}

\begin{figure}
\centering
\includegraphics[width=0.98\columnwidth]{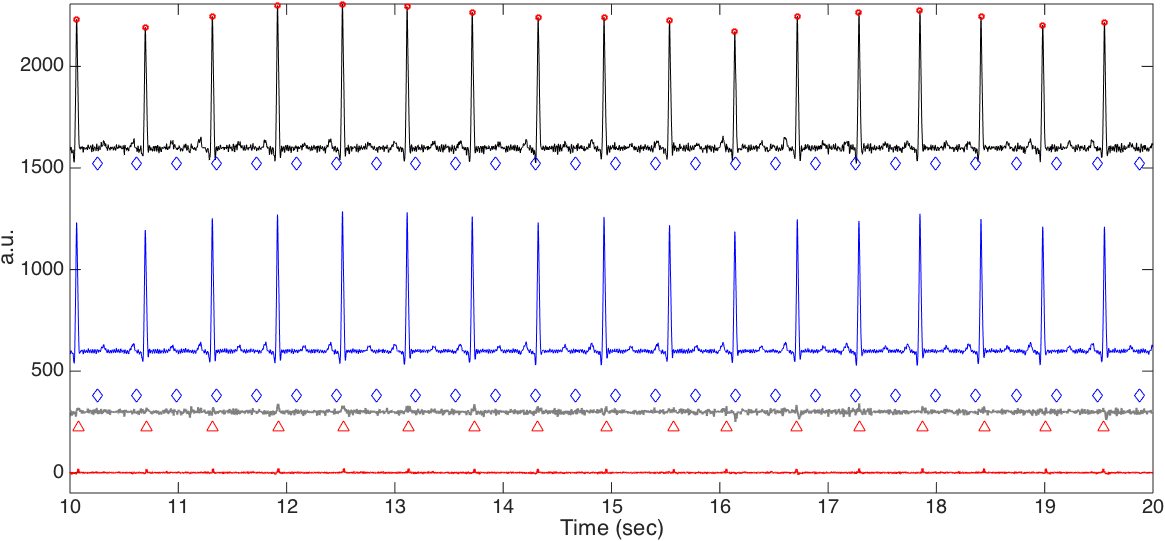}\\
\includegraphics[width=0.98\columnwidth]{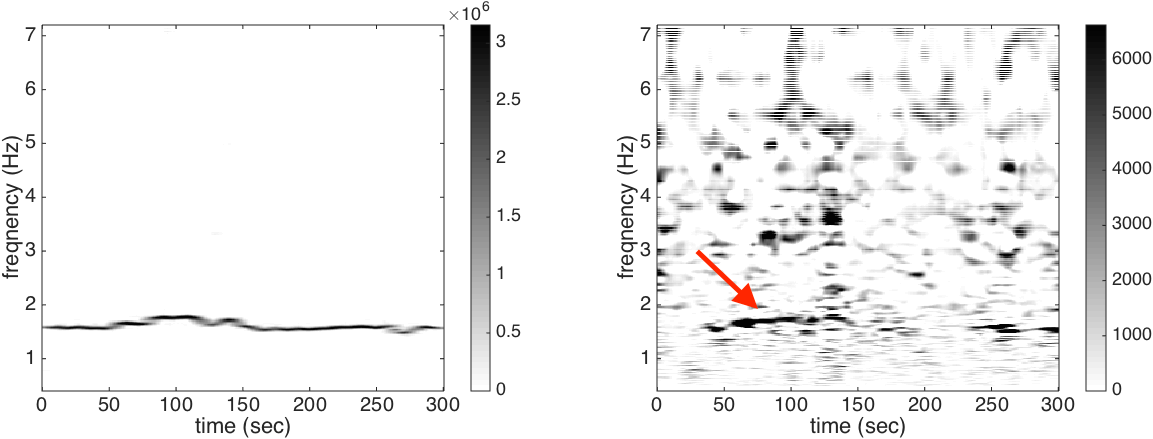}
\caption{The result of channel 2 of a59 in the CinC2013 database. In the top subplot, all relevant signals are shown together for a visual comparison. The clean maECG signal is shown in black (shifted up by 1600 units) with the detected maternal R peaks in round red circles and the annotated fetal R peaks in blue diamonds on the top row; the estimated maECG signal is shown in blue  (shifted up by 600 units) on the second row; the rough fetal ECG is shown in dark gray (shifted up by 300 units) with the detected R peaks in red upper triangles and the annotated R peaks in blue diamonds on the third row; the fECG estimation is shown in red on the bottom row. It is clear that the fetal R peaks are all detected incorrectly.
In the bottom subplot, the de-shape STFT of the maECG is shown on the left and the de-shape STFT of the rough fECG is shown on the right. The red arrow indicated the ``dominant curve'' in the de-shape STFT, which comes from the incomplete removal of the maECG signal.
}
\label{fig:CinC2013a59part1}
\end{figure}

Our proposed algorithm falls in the ``iterative'' category -- estimate and remove the maternal component first, and get the fIHR and fECG from the left.
We do have the accumulated error issue when we estimate the fECG, and we count on the median filter to alleviate this error. One natural question is to ask if it is possible to generate a single-lead fECG algorithm in the joint category, in order to alleviate this problem, and the answer is positive.  
Precisely, the mIHR and fIHR could be estimated simultaneously by the de-shape STFT, as is shown in Figure \ref{fig:real_deSTFT}. The estimated R peaks of each component could then be applied to estimate the maECG and fECG by the nonlocal median algorithm.
In brief, we could consider an algorithm falling in the joint category, which is composed of two steps:
(1) jointly estimate the mIHR and the fIHR by the de-shape STFT; that is, carry out Optional step 1 in the algorithm;
(2) jointly approximate the maECG and fECG signal by the nonlocal mean.
However, a preliminary study showed that without a more sophisticated curve extraction algorithm, this approach does not outperform what we propose in the current paper. Its potential will be explored in the future work.

\subsection{Several theoretical and algorithmic topics}

We should discuss more about each step in the proposed algorithm. First of all, notice that the proposed algorithm is specifically designed to handle a nonstationary signal, which is commonly encountered in physiology. The de-shape STFT respects the local information hidden in a nonstationary time series, and decouple the non-sinusoidal oscillatory pattern from the IF; the nonlocal mean, while it is not local, takes the nonlinear relationship between the phase and the oscillation (the nonlinear relationship between the RR interval and QT interval) into account. Thus, the proposed algorithm could be applied to analyze a long time series without modification, and the hard truncation of the long time series into pieces is not needed.

We have seen the potential of extracting more information hidden inside the single-lead aECG signal by the ANH model and the de-shape STFT. In the past few years, different ideas in the TF analysis field have been proposed, regarding the model and the algorithm, and it is possible to incorporate them into the current algorithm to achieve a better practical result. For example, the Blaschke decomposition and unwinding based on the analytic signal analysis technique \cite{Coifman_Steinerberger:2015}, the synchrosqueezing transform, or the concentration of frequency and time (ConceFT) \cite{Daubechies_Wang_Wu:2016}. However, to simplify the discussion, we do not carry out this combination in this paper. While de-shape STFT works well, we cannot ignore a major limitation in the curve extraction step for the IHR estimation. Precisely, when the energy of the fECG signal is greater then that of the maECG, then the curve extraction algorithm could easily go astray; that is, when the fECG has an almost equal or larger energy than the maECG, we could see two dominant curves, which may or may not intersect, and the curve extraction algorithm could get confused. While we could apply the physiological constraint to distinguish which curve belongs to the maECG, for example, the fIHR is in general faster than mIHR (Optional step 3), it is not universal, and will limit the method to normal subjects. We thus need more tools to handle this issue.

Next, we mention that the ability to obtain an accurate IHR estimation is the key step to an accurate R peak location estimation by the beat tracking algorithm. The technical limitation in obtaining a good IF information might explain why the intuitive and powerful beat tracking technique is not widely applied up to now, although it was introduced in the music signal analysis field long time ago \cite{ellis2007beat}. In sum, the combination of the de-shape STFT, or any other technique providing an accurate IF estimation, is essential.
Also note that this combination could offer an alternative way to estimate the R peaks or other fiducial points, and even to other signal processing problems. %, in general to estimate the time-varying wave-shape function, in addition to the detection and separation problems.

Furthermore, to the best of our knowledge, this work is the first one combining the nonlocal median and the nonlinear manifold model to analyze the biomedical time-series analysis, and particularly reconstructing the time-varying wave-shape function. Taking the nonlinear manifold structure to analyze a time series is certainly not a new idea \cite{Richter1998,KarvounisTsipouras2009,Kotas2010}. However, in the past, the focus was on decoupling maECG and fECG by the locally linear projection on the sequential beats. In this work, the nonlinear geometric structure is further explored to apply the nonlocal median. In general, we could apply other manifold learning techniques to further improve the algorithm.
From a data analysis viewpoint, it is well known that a manifold structure might still be too restrictive to be the best model for this dataset, while it could be low dimensional and nonlinear with nontrivial topological structures, and could serve as a good mathematical framework to understand different algorithms. Thus, finding a more flexible model to include more physiological facts is a critical issue in the future study.
While the $L^1$ norm is associated with the nonlocal median algorithm, to enhance the sparsity feature, it is possible to consider the $L^p$ norm, where $0<p<1$, which has been applied in the recently proposed nonlocal patch regression for image denoising \cite{ChaudhurySinger2013}.

\begin{figure}
\centering
\includegraphics[width=0.98\columnwidth]{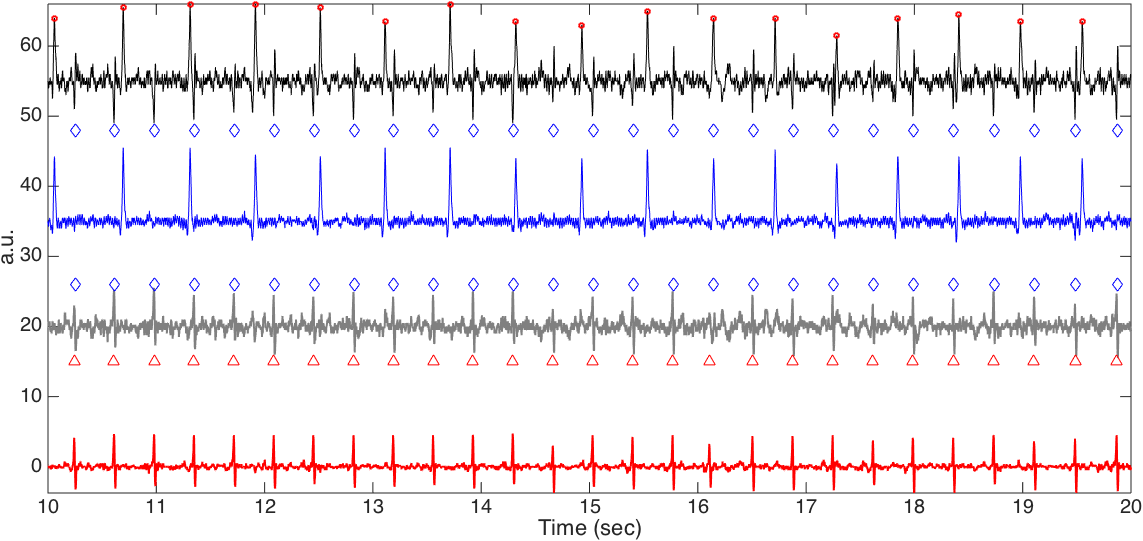}\\
\includegraphics[width=0.98\columnwidth]{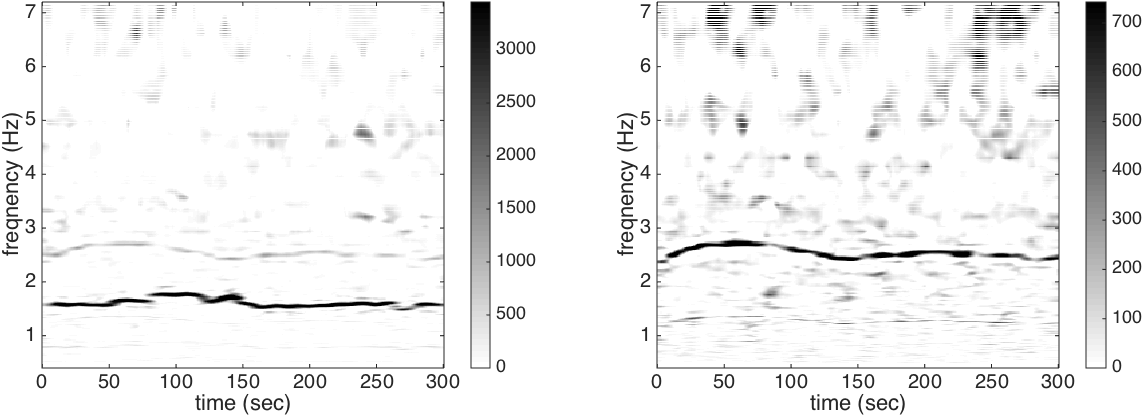}
\caption{The result of the difference between channel 2 and channel 3 of a59 in the CinC2013 database. In the top subplot, all relevant signals are shown together for a visual comparison. The clean maECG signal is shown in black (shifted up by 55 units) with the detected maternal R peaks in round red circles and the annotated fetal R peaks in blue diamonds on the top row; the estimated maECG signal is shown in blue  (shifted up by 35 units) on the second row; the rough fetal ECG is shown in dark gray  (shifted up by 20 units) with the detected fetal R peaks in round red circles and the annotated fetal R peaks on red upper triangles on the third row; the fECG estimation is shown in red on the bottom row.
In the bottom subplot, the de-shape STFT of the maECG is shown on the left and the de-shape STFT of the rough fECG is shown on the right. Clearly, the fetal R peaks is perfectly recovered.
}
\label{fig:CinC2013a59part2}
\end{figure}

We have extensively extracted available information hidden in the single-lead aECG signal. When there are more than one channel, it could be beneficial to combine results from different channels, or to integrate the proposed algorithm with existing multiple-lead algorithms to further improve the accuracy. The benefit of doing so has been shown with the case a59 in the CinC2013 database in Figure \ref{fig:CinC2013a59part2}. Another direct benefit of using multiple-lead algorithms is guiding the mIHR estimation by the curve extraction, when the fECG has a stronger energy than the maECG. In practice, we could also expect to combine other kinds of signals, like the contact photoplethysmogram signal commonly equipped in modern wearable devices, to guide the mIHR estimation. The result of this kind of combination and the analysis of multiple channel signals will be reported in the future work.

\subsection{Several practical topics}

In this study, we do not carry out any optimization to choose the best parameters, but just choose them in an ad hoc fashion. For example, due to the inevitable Heisenberg uncertainty principal, in general the window should be long enough to capture enough spectrum information, but it could not be too long, otherwise the local information could be missed. In practice we found that a window that could cover about 5-10 oscillations is a good choice. On the other hand, when the signal is noisy, we could get more stable results if the window is longer. Based on the physiological constraint and the discussion, we thus choose a 10 sec long window for the very noisy simulated signal, and a 5 sec window for the less noisy real databases. We mention that the algorithm in general is stable to the choice of parameters, and the performance of the algorithm could be further improved by running a systematic parameter optimization for a specific mission.

The proposed fECG extraction algorithm is just one among many successful algorithms. Each algorithm has its own strength and weakness, and it is generally believed that there might not exist a single algorithm that could handle all different situations and different signals. Thus, it would be of great interest to consider the possibility of combining the proposed algorithm with other methods, like the combination proposed in \cite{BeharOsterClifford2014}. More practically, while in the real life the noise could be highly nonstationary, we may want to combine the notion of signal quality index (SQI) to help guide us in selecting the ``good'' signal for analysis \cite{Behar2014,LiuLiDiMaria2014,Gambarotta2016}. In general, if the noise is really too big and bypass the information limitation, then there is no hope to recover anything. However, if the noise is not that big, then some denoising techniques could certainly help. In this work, to simplify the discussion, we do not consider any denoising technique before running de-shape STFT or beat tracking. However, it could be of great help if we could incorporate a good de-noise scheme into the algorithm. All these interesting practical issues will be explored in the near future.

\subsection{Several clinical topics}

\begin{figure}
\centering
\includegraphics[width=0.5\columnwidth]{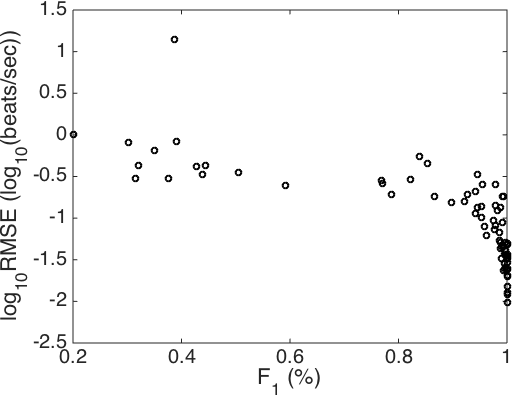}
\caption{The RMSE and $F_1$ of the 75 subjects in the CinC2013 database is plotted.
}
\label{Figure:F1vsHRV0}
\end{figure}

In clinics, the fECG signal could provide at least two different kinds of physiological information -- the HRV and the electrophysiological dynamics. For the HRV analysis, the most important ingredient is having an accurate R peak detection algorithm. One of the main strength of the propose algorithm is providing the state-of-the-art MAE in the field. For example, in the real database, the averaged MAE is about 5 msec, which is roughly equivalent to the R peak information from an fECG signal sampled at about 200Hz, and its clinical value does deserve a further evaluation. However, note that when we evaluate the MAE, we focus only on the TP beats, which means that we could only get a good HRV analysis when $F_1$ is high enough. See Figure \ref{Figure:F1vsHRV0} for an illustration of the relationship between the HRV analysis and the $F_1$. This figure is generated in the following way.
For each case in CinC2013 (75 in total), we evaluate the root mean square error (RMSE) to quantify how accurate we could recover the fIHR for the HRV analysis. Based on the Task Force \cite{TaskForce:1996}, we get the fIHR's from the estimated fetal R peaks and from the annotated fetal R peaks, and RMSE between these two fIHR is evaluated. Note that we take all estimated fetal R peaks to simulate the real scenario. We could see that when $F_1$ is slightly worse than 95\%, the RMSE is as high as 0.1 beats/sec. This preliminary result emphasizes the importance of getting an accurate $F_1$.
An extensive study of this important topic will be reported in the future work.

For the fECG morphology reconstruction, although the result is convincing, note that we could not infer too much electrophysiological information from the single-lead fECG signal, even if the reconstruction is perfect. Indeed, due to the variation of relative cardiac axes between the mother and fetus, the projection direction of fECG is in general unknown, even if the abdominal lead system on the abdominal surface is standardized. To explore this direction, we need to design or choose the optimal single abdominal lead location under the complicated time-varying dynamics, which is adaptive to the mother, or we may take multiple leads and proceed with an adaptive way to reconstruct the VCG signal.

Uterine contraction is in general a difficult problem and big challenge in the fECG signal analysis, since it behaves like a huge noise in the aECG signal.
In general, to handle such a huge noise, some a priori knowledge of the noise is needed. While the current method could provide a reasonable result in Case 3 of the simulated database, it does not have the ability to handle the uterine contraction, and that is why we always get an $F_1$ less then 80\% in the 5 min result.
While it is out of the scope of this paper, we mention that one possible approach is to incorporate the physiological behaviors underlying cardiac signals, a realistic model of an individual fetus' ECG, or the statistical behavior of the noise associated with the uterine contraction to improve the result \cite{Clifford2005,SameniClifford2007,SameniShamsollaha2007}.

It is not a reasonable assumption that the mother or the fetus are both healthy in the real life. Sometimes we might encounter subjects with a problematic heart. In Case 4 of the simulated database, the situation when the fetus and mother both have frequent ectopic beats, like ventricular pre-contraction (VPC), is considered.
We see that the nonlocal median algorithm gives us a reasonable cardiac activity estimation. Indeed, the distance between an ectopic beat and a normal beat is in general larger than the distance between two ectopic beats and two normal betas. Therefore, the nonlocal nature of the nonlocal median algorithm could accurately estimate the cardiac activity, since VPC and normal beats are clustered into two groups. Note that if VPC is polymorphic and/or the VPC number is small, the problem becomes more challenging and a different approach is needed.

The twin pregnancy problem was relatively rarely discussed, except in \cite{Taylor2003,Niknazar2013}. In Case 5 of the simulated database, although we do not specifically study in this paper, we mention that the proposed algorithm has the potential to handle even multiple pregnancy problem. Indeed, the mIHR and fIHR's of different fetuses could be simultaneously revealed in the de-shape STFT. However, it is still challenging to utilize this single channel information at this stage, since there is no a priori information about the relationship between the IHR's and signal strengths of those fetuses. This limits the curve extraction step in the de-shape STFT, as the IHR's and energy of two fetus might be similar. Thus, although we could see all the fIHR's, to extract the fIHR from the TF representation based on the current curve extraction algorithm is challenging, and we need a more sophisticated curve extraction algorithm to do it.
The study of the above-mentioned clinical topics will be explored in the future work.

\subsection{Limitation}

The discussion could not be complete without mentioning the limitation.
While the strength of the proposed algorithm is confirmed on one simulation and two real databases, a larger scale clinical study is needed to further confirm the usefulness of the proposed algorithm since the size of the publicly available databases, \textit{adfecgdb} and \textit{CinC2013}, is small.
Second, although {\em fecgsyn} database provides a universal comparison platform with many interesting examples, the model at this stage is still oversimplified, and that might explain the high accuracy of our proposed algorithm. Many other limitations have been well discussed in \cite{Andreotti2016}. A better ``ground truth'' should be considered for the evaluation purpose. For example, a well established sheep model \cite{10.3389/fped.2014.00038} could provide a gold standard test bed for the algorithm.

\section{Conclusion}

In this paper, we propose a novel fECG extraction algorithm depending mainly on a single-lead aECG based on a nonlinear TF analysis technique, de-shape STFT, and the nonlocal median. In addition to providing the theoretical model and mathematical guarantee, the algorithm is evaluated on a simulated fECG database and tested on two real databases with experts' annotation. The main novelty in our algorithm is an extensive utilization of hidden information inside the aECG signal, that is, both the frequency and energy information and the nonlinear relationship between consecutive cardiac activities. In addition to solving clinical problems, we mention that the combination of the de-shape STFT and nonlocal median has a potential to deal with more general multi-component oscillatory signals in other fields.

\section{Acknowledgement}
Hau-tieng Wu's research is partially supported by Sloan Research Fellow FR-2015-65363. Part of this work was done during Hau-tieng Wu's visit to National Center for Theoretical Sciences, Taiwan, and he would like to thank NCTS for its hospitality. Hau-tieng Wu acknowledges Professor Ingrid Daubechies for the continuous discussion about various topics and Professor Martin Frasch for the discussion about clinical needs.

\bibliographystyle{amsplain}
\bibliography{fECGCepstrum}

\clearpage

\begin{table}[!ht]
\centering
\caption{Results of fetal R peaks estimation -- median $F_1$ and MAE and their IQRs for different cases and noise levels among 10 subjects and 5 rounds. The 1min result is reported in the following way. For each subject, noise level, case, and round, we take the channel and the 1-minute subset out of the 5-minute signal that leads to the highest $F_1$ to report the result. Then, for each noise level and each case, the median and IQR of all subjects and rounds are shown as the final result, as is suggested in \cite{Andreotti2016}. For the 5min result, for each subject, noise level, case, and round, we take the channel that leads to the highest $F_1$, and then for each noise level and each case, we report the median and IQR of all subjects and rounds.}
\begin{tabular}{|l|c|ccccc|}
\hline
  \multicolumn{2}{|c|}{} & Case 0 & Case 1 & Case 2 & Case 3 & Case 4 \\
\hline
\hline
\multicolumn{2}{|c|}{} &\multicolumn{5}{c|}{$F_1$ (\%)} \\
\hline
\multirow{2}{*}{12 dB} &1min& 100 (0) & 100 (0) & 100 (0) & 100 (0) & 100 (0) \\
&5min&99.93 (0.38) & 99.92 (0.21) & 100 (0.87) & 72.96 (10.32) & 99.85 (0.47)\\
\hline
\multirow{2}{*}{9 dB} & 1min& 100 (0) & 100 (0) & 100 (0) & 100 (0) & 100 (0) \\
&5min& 99.9 (0.61) & 99.91 (0.29) & 99.93 (0.27) & 73.6 (9.44) & 99.58 (3.75)\\
\hline
\multirow{2}{*}{6 dB} &1min &100 (0.73) & 100 (0.88) & 100 (0.67) & 100 (0.43) & 100 (7.83)\\
&5min& 99.89 (1.38) & 99.82 (11.01) & 99.91 (9.56) & 73.72 (13.58) & 99.31 (22.52)\\
\hline
\multirow{2}{*}{3 dB} &1min &100 (71.53) & 100 (0.88) & 100 (69.14) & 100 (6.11) & 100 (19.82)\\
&5min& 99.12 (76.49) & 98.72 (6.17) & 99.44 (78.63) & 72.11 (28.53) & 98.6 (45.75)\\
\hline
\multirow{2}{*}{0 dB} &1min &95.63 (56.14) & 97.99 (19.31) & 97.9 (64.75) & 99.61 (37) & 93.76 (64.04)\\
&5min& 88.24 (74.69) & 83.46 (38.1) & 81.17 (73.67) & 64.25 (38.3) & 85.84 (74.18)\\
\hline
\hline
\multicolumn{2}{|c|}{} &\multicolumn{5}{c|}{MAE (msec)} \\
\hline
\multirow{2}{*}{12 dB} & 1min&0.85 (1.54) & 0.5 (0.77) & 0.68 (1.2) & 0.53 (0.52) & 2.15 (1.46)\\
& 5min& 0.82 (1.82) & 0.58 (0.78) & 0.63 (1.59) & 3.07 (1.54) & 2.32 (1.63) \\
\hline
\multirow{2}{*}{9 dB} & 1min&1.01 (2.83) & 0.75 (0.65) & 0.7 (0.49) & 0.51 (0.58) & 3.21 (1.77)\\
&5min &0.96 (2.96) & 0.71 (0.7) & 0.77 (0.37) & 3.15 (1.62) & 3.2 (1.77) \\
\hline
\multirow{2}{*}{6 dB} &1min &1.21 (3.04) & 0.92 (2.7) & 1.06 (2.16) & 0.85 (1.01) & 4.03 (5.56) \\
& 5min&1.15 (3.17) & 0.93 (2.48) & 1 (3.11) & 3.43 (1.99) & 4.01 (5.52) \\
\hline
\multirow{2}{*}{3 dB} & 1min& 2.23 (10.64) & 1.43 (3.18) & 1.41 (4.15) & 1.11 (6.67) & 5.13 (11.54)\\
& 5min&1.9 (12.18) & 1.76 (4.16) & 1.38 (4.13) & 3.77 (8.61) & 5.07 (11.89)\\
\hline
\multirow{2}{*}{0 dB} &1min & 5.65 (8.06) & 4.36 (8.4) & 3.5 (13.46) & 2.13 (11.66) & 6.6 (13.47) \\
&5min&5.27 (7.85) & 5.59 (9.47) & 4.56 (10.93) & 5.53 (11.47) & 6.26 (13.34)\\
\hline
\end{tabular}
\label{tab: fecgsynF1}
\end{table}

\begin{table}[!ht]
\centering
\caption{Results of fetal R peaks estimation for the \textit{tenth} best channel -- median $F_1$ and MAE and their IQRs for different cases and noise levels among 10 subjects and 5 rounds. The 1min result is reported in the following way. For each subject, noise level, case, and round, we take the channel and the 1-minute subset out of the 5-minute signal that leads to the \textit{tenth} highest $F_1$ to report the result. Then, for each noise level and each case, the median and IQR of all subjects and rounds are shown as the final result, as is suggested in \cite{Andreotti2016}. For the 5min result, for each subject, noise level, case, and round, we take the channel that leads to the \textit{tenth} highest $F_1$, and then for each noise level and each case, we report the median and IQR of all subjects and rounds.}
\begin{tabular}{|l|c|ccccc|}
\hline
  \multicolumn{2}{|c|}{} & Case 0 & Case 1 & Case 2 & Case 3 & Case 4 \\
\hline
\hline
\multicolumn{2}{|c|}{} &\multicolumn{5}{c|}{$F_1$ (\%)} \\
\hline
\multirow{2}{*}{12 dB} &1min& 100 (0) & 100 (0) & 100 (0.36) & 100 (0.4) & 100 (1.74)\\
&5min&99.82 (9.28) & 99.88 (0.35) & 99.39 (14.41) & 64.17 (19.45) & 99.02 (16.1)\\
\hline
\multirow{2}{*}{9 dB} & 1min&100 (1.25) & 100 (74.63) & 100 (4.62) & 100 (2.08) & 99.2 (3.13) \\
&5min& 99.48 (5.48) & 99.53 (78.69) & 99.48 (27.05) & 64.61 (17.64) & 97.71 (10.31)
\\
\hline
\multirow{2}{*}{6 dB} &1min &100 (10.37) & 100 (1.64) & 99.23 (64.58) & 99.6 (14.44) & 98.94 (12.78)\\
&5min& 96.02 (43.21) & 98.06 (6.49) & 94.48 (74.91) & 60.57 (21.72) & 93.71 (28.81)\\
\hline
\multirow{2}{*}{3 dB} &1min &97.89 (59.78) & 99.18 (9.54) & 99.8 (80.33) & 99.7 (66.55) & 94.79 (29.41)\\
&5min& 87.54 (74.19) & 95.27 (21.42) & 78.4 (90.88) & 60.31 (47.41) & 87.48 (48.56)\\
\hline
\multirow{2}{*}{0 dB} &1min &51.55 (71.07) & 80.53 (80.79) & 46.34 (67.88) & 83.73 (68.65) & 83.69 (38.35)\\
&5min& 43.7 (69.08) & 49.29 (90.35) & 32.07 (49.34) & 45.65 (41.85) & 66.02 (57.91)
\\
\hline
\hline
\multicolumn{2}{|c|}{} &\multicolumn{5}{c|}{MAE (msec)} \\
\hline
\multirow{2}{*}{12 dB} & 1min&1.09 (1.8) & 0.84 (1.23) & 1.29 (2.66) & 1.03 (2.26) & 3.87 (2.97)\\
& 5min&1.16 (2.14) & 0.75 (1.07) & 1.17 (2.61) & 4.41 (3.72) & 3.86 (2.94) \\
\hline
\multirow{2}{*}{9 dB} & 1min&1.36 (5.53) & 0.72 (1.5) & 1.37 (5.99) & 1.51 (2.31) & 4.74 (3.76)\\
&5min &1.34 (6.02) & 0.64 (1.71) & 1.48 (5.58) & 5.18 (3.51) & 4.74 (4.15)\\
\hline
\multirow{2}{*}{6 dB} &1min &2.87 (7.36) & 1.36 (3.68) & 4.4 (13.72) & 1.36 (5.73) & 6.75 (5.03) \\
& 5min&3.17 (7.5) & 2.52 (4.39) & 4.33 (17.84) & 5.47 (5.92) & 6.74 (6.27)\\
\hline
\multirow{2}{*}{3 dB} & 1min& 4.44 (11.48) & 2.98 (4.53) & 2.18 (13.44) & 2.09 (19.65) & 8.13 (8.8)\\
& 5min&6 (11.63) & 3.54 (5.34) & 3.14 (14.91) & 5.49 (15.82) & 8.8 (8.56) \\
\hline
\multirow{2}{*}{0 dB} &1min &11.46 (12.85) & 6.18 (16.45) & 14.41 (12.98) & 8.01 (19.81) & 9.06 (9.89)\\
&5min&13.29 (13.67) & 6.22 (17.23) & 15.57 (13.19) & 12.11 (16.83) & 9.04 (10.82)\\
\hline
\end{tabular}
\label{tab: fecgsynF1part2}
\end{table}

\begin{table}[!ht]
\centering
\caption{Results of fECG waveform estimation over the 5-minute signal -- median correlation and its IQR for different cases and noise levels among 10 subjects and 5 rounds.
For each subject, noise level, case, and round, we take the channel that leads to the highest $F_1$ (and the \textit{thenth} highest $F_1$), and then for each noise level and each case, we report the median and IQR of all subjects and rounds.}
\begin{tabular}{|l|c|ccccc|}
\hline
\multicolumn{2}{|c|}{}   & Case 0 & Case 1 & Case 2 & Case 3 & Case 4 \\
\hline
\hline
\multicolumn{2}{|c|}{} &\multicolumn{5}{c|}{Correlation} \\
\hline
\multirow{2}{*}{12 dB} &highest& 0.962 (0.067) & 0.977 (0.044) & 0.984 (0.052) & 0.983 (0.017) & 0.97 (0.032) \\
&10th highest&0.953 (0.097) & 0.96 (0.047) & 0.959 (0.062) & 0.973 (0.052) & 0.963 (0.108) \\
\multirow{2}{*}{9 dB} &highest& 0.962 (0.076) & 0.968 (0.05) & 0.977 (0.021) & 0.978 (0.034) & 0.962 (0.039) \\
&10th highest&0.93 (0.222) & 0.943 (0.753) & 0.953 (0.12) & 0.97 (0.075) & 0.942 (0.164) \\
\multirow{2}{*}{6 dB} &highest& 0.954 (0.076) & 0.958 (0.083) & 0.949 (0.069) & 0.977 (0.053) & 0.957 (0.141) \\
&10th highest&0.912 (0.256) & 0.919 (0.108) & 0.897 (0.374) & 0.948 (0.118) & 0.926 (0.191) \\
\multirow{2}{*}{3 dB} &highest& 0.919 (0.48) & 0.94 (0.103) & 0.939 (0.276) & 0.965 (0.133) & 0.95 (0.216) \\
&10th highest&0.859 (0.348) & 0.925 (0.29) & 0.923 (0.428) & 0.945 (0.47) & 0.863 (0.37) \\
\multirow{2}{*}{0 dB} &highest& 0.892 (0.404) & 0.894 (0.289) & 0.9 (0.371) & 0.942 (0.215) & 0.875 (0.386) \\
&10th highest&0.656 (0.378) & 0.846 (0.518) & 0.768 (0.3) & 0.886 (0.345) & 0.778 (0.375) \\
\hline
%\hline
%\multicolumn{2}{|c|}{} &\multicolumn{5}{c|}{$L^\infty$ (mV)} \\
%\hline
%\multirow{2}{*}{12 dB} & highest&0.14 (0.17) & 0.16 (0.1) & 0.1 (0.06) & 0.12 (0.1) & 0.07 (0.04) \\
%&10th highest&0.11 (0.1) & 0.1 (0.08) & 0.08 (0.06) & 0.07 (0.09) & 0.07 (0.05) \\
%\multirow{2}{*}{9 dB} & highest&0.13 (0.09) & 0.15 (0.15) & 0.10 (0.08) & 0.13 (0.10) & 0.11 (0.09)\\
%&10th highest&0.13 (0.13) & 0.27 (0.81) & 0.11 (0.07) & 0.07 (0.05) & 0.12 (0.09) \\
%\multirow{2}{*}{6 dB} & highest&0.15 (0.08) & 0.17 (0.20) & 0.13 (0.12) & 0.13 (0.09) & 0.11 (0.09) \\
%&10th highest&0.12 (0.08) & 0.13 (0.08) & 0.16 (0.12) & 0.11 (0.09) & 0.08 (0.07) \\
%\multirow{2}{*}{3 dB} & highest&0.18 (0.16) & 0.19 (0.16) & 0.15 (0.09) & 0.14 (0.11) & 0.16 (0.14) \\
%&10th highest&0.15 (0.12) & 0.27 (0.22) & 0.17 (0.15) & 0.13 (0.1) & 0.13 (0.08) \\
%\multirow{2}{*}{0 dB} & highest&0.20 (0.10) & 0.26 (0.19) & 0.17 (0.09) & 0.16 (0.12) & 0.24 (0.18) \\
%&10th highest&0.22 (0.21) & 0.3 (0.25) & 0.18 (0.13) & 0.18 (0.19) & 0.22 (0.12) \\
%\hline
\end{tabular}
\label{tab: fecgsynMorphology}
\end{table}

\begin{table}
\centering
\caption{$F_1$ score, MAE, PPV, and SE of all channels and all subjects over the whole $5$ minute signals in the adfecgdb database. For each subject, the best result among 4 channels is marked in the boldface.}
\begin{tabular}{|c|c|cccc|}
\hline
Subject & Channel & $F_1$ (\%) & MAE (msec) & PPV (\%) & SE (\%) \\
\hline
\hline
\multirow{4}{*}{r01} &	1&	\textbf{99.53} &\textbf{1.5227} &\textbf{99.38} &\textbf{99.69}\\
&	2&	75.21 &4.4004 &72.33 &78.32\\
&	3&	87.35 &4.4046 &86.41 &88.30\\
&	4&	78.83 &4.9358 &77.53 &80.19\\
\hline
\multirow{4}{*}{r04}&	1&	82.15 &9.0635 &81.63 &82.67\\
&	2&	\textbf{98.81} &7.6447 &\textbf{98.73} &\textbf{98.89}\\
&	3&	97.54 &7.2622 &97.46 &97.62\\
&	4&	98.41 &\textbf{6.9065} &98.26 &98.57\\
\hline
\multirow{4}{*}{r07}&	1&	84.81 &10.5816 &84.20 &85.42\\
&	2&	98.48 &8.3182 &98.25 &98.72\\
&	3&	\textbf{99.44} &7.6329 &\textbf{99.36} &\textbf{99.52}\\
&	4&	99.28 &\textbf{7.3048} &99.20 &99.36\\
\hline
\multirow{4}{*}{r08}&	1&	\textbf{97.85} &\textbf{2.0219} &\textbf{97.26} &\textbf{98.46}\\
&	2&	84.44 &3.7191 &82.69 &86.27\\
&	3&	71.42 &5.3655 &69.49 &73.46\\
&	4&	69.69 &5.2397 &71.92 &67.59\\
\hline
\multirow{4}{*}{r10}&	1&	\textbf{98.65} &\textbf{2.8119} &\textbf{98.26} &\textbf{99.04}\\
&	2&	98.49 &3.5919 &98.26 &98.73\\
&	3&	78.52 &8.5177 &76.28 &80.89\\
&	4&	95.25 &4.1681 &94.79 &95.70\\
\hline
\end{tabular}
\label{tab: adfecgdbF1}
\end{table}

\clearpage

\end{document}